\documentclass[aip,pop,reprint]{revtex4-2}

\usepackage{hyperref}
\usepackage[utf8]{inputenc}
\usepackage[english]{babel}
\usepackage{bm}
\usepackage{amsmath}
\usepackage{amssymb}
\usepackage[shortcuts]{extdash}
\usepackage{graphicx}
\usepackage{tabularx}
\usepackage{soul,xcolor}
\DeclareUnicodeCharacter{2010}{-}

\renewcommand{\vec}[1]{\mathrm{\mathbf{#1}}}

\begin{document}
\DeclareGraphicsExtensions{.pdf,.png,.jpg}

\title{Waveshape of THz radiation produced by two-color laser-induced air plasmas}

\author{A. Stathopulos}
\affiliation{CEA, DAM, DIF, 91297 Arpajon, France}
\affiliation{Universit\'e Paris-Saclay, CEA, LMCE, 91680 Bruy\`eres-le-Ch\^atel, France}
\author{S. Skupin} 
\affiliation{Institut Lumière Matière, UMR 5306 - CNRS, Université de Lyon 1, 69622 Villeurbanne, France}
\author{B. Zhou} 
\affiliation{DTU Electro - Department of Electrical and Photonics Engineering, Technical University of Denmark,
2800 Kongens Lyngby, Denmark}
\author{P. U. Jepsen} 
\affiliation{DTU Electro - Department of Electrical and Photonics Engineering, Technical University of Denmark,
2800 Kongens Lyngby, Denmark}
\author{L. Berg\'e}
\email{luc.berge@u-bordeaux.fr}
\affiliation{Centre des Lasers Intenses et Applications, Université de Bordeaux-CNRS-CEA, 33405 Talence Cedex, France}

\date{\today}

\begin{abstract}
The spatial and spectral distributions of terahertz (THz) pulses emitted by two-color air plasmas are theoretically investigated for focused laser pulses and in the filamentation regime. We derive a so-called ''augmented'' conical emission model, which, similarly to the one originally proposed by You et al.\ [Phys.\ Rev.\ Lett.\ {\bf 109}, 183902 (2012)], involves phase matching between laser harmonics along the plasma channel, the plasma density and length, and the emitted frequency as key parameters. Our augmented model, however, treats envelope effects and accounts for transverse variations of both plasma- and Kerr-driven potential THz emitters. We highlight the importance of the characteristic spatio-spectral distributions of these two conversion mechanisms in the expression of the angular radiated power. The results of our model are successfully compared with data provided by a comprehensive, fully space and time-resolved unidirectional solver. Importantly, these numerical simulations clear up the effective plasma length along which THz emission develops, compared with the dephasing length along which the laser fundamental and second harmonic become out-of-phase. The impact of common optical aberrations, such as sphericity, astigmatism, and coma, on the THz generation is also investigated. Aberrations are shown to generally decrease the laser-to-THz conversion efficiency and potentially induce spatial asymmetries and narrowing in the THz spectra. 
\end{abstract}

\maketitle

\section{Introduction} 
The production of terahertz (THz) radiation is receiving increasing interest in a wide range of domains, such as homeland security \cite{Tonouchi, Currie, Federici}, medical diagnosis \cite{Woodward2001, Woodward2003, Pickwell}, art conservation \cite{Fukunaga, Yasui, Dandolo2015a, Dandolo2015b} or archeology \cite{Bianca, Yin}. THz waves can propagate through a few millimeters of optically opaque materials, provided the water concentration is low \cite{Thrane, Ronne}, and can be used to identify complex molecules by coherent THz spectroscopy \cite{Fattinger_1988, Fattinger_1989, Walther2000, Neu}. Other notable applications of THz waves include control of leaf water content \cite{Nuss1995, Gente2015, Gente2018}, plasma excitation  \cite{Curcio}, electron acceleration \cite{Zhang2018}, high harmonic (sideband) generation \cite{Zaks2012, Schubert2014, Langer2016}, or the fabrication of ferromagnetic structures \cite{Shalaby}. Several emitters can nowadays produce intense pulsed or continuous-wave THz radiation, including photoconductive antennas \cite{Burford, Park}, molecular (CH$_3$OH) gas sources \cite{Zhong}, quantum cascade lasers \cite{Kohler, Williams, Li} or $\chi^{(2)}$ crystals serving for optical rectification \cite{Vicario2014}. However, THz time-domain spectroscopy (THz-TDS) requires broadband THz radiation to access extended absorption spectra, allowing the identification of molecules or precise determination of the physics of broadband processes such as relaxation and conductivity. This need can be fulfilled by ionizing a gas with an ultrashort multi-color laser pulse. This method generates strong ($\sim 0.1 - 1$ GV/m) and ultra-broadband (exceeding 100 THz) terahertz fields with significant conversion efficiencies ($> 10^{-4}$) \cite{Cook, Koulouklidis}. Another benefit of this technique is the non-destructive nature of the mechanisms at play, e.g., in air plasmas. The laser driver usually comprises only two harmonics. However, recent theoretical and experimental results showed that adding more harmonics to the optical beam can dramatically increase the THz yield \cite{Sawtooth, SawtoothCPS, Liu3c, Vaicaitis3c}.

When a temporally asymmetric femtosecond light pulse composed of a fundamental component and its second harmonic reaches the $\sim 10^{14}$ W/cm$^2$ intensity range, sufficient to photoionize a gas (here, ambient air) in the tunnel regime \cite{Keldysh} an underdense plasma is created. The electrons freed by this process are then accelerated by the asymmetric laser field and induce a photocurrent that contains a low-frequency component. This quasi-DC contribution causes the plasma to radiate in the THz domain. Although this process is usually performed in the so-called focused regime, where the plasma spot is formed a few tens of cm after a focusing element, recent works have confirmed its efficiency also in the long-range filamentation regime \cite{Daigle2012} that results from a sequence of local equilibria between Kerr self-focusing and plasma defocusing. This regime can lead to meter-long filaments \cite{ROP2007} and could be advantageous for THz-TDS of explosives for safety reasons \cite{EPL_Altesse,Talbi2023}.

One specificity of THz wave generation by two-color air plasmas is the typically conical shape of the THz radiation \cite{Andreeva}. In their seminal work, You et al.~\cite{You2012,You2013} advocated that conical emission is due to interferences between the THz waves produced at different places along a generated cylindrical homogeneous plasma, as their polarity changes with the phase angle between the two laser harmonics caused by optical and plasma dispersion. Importantly, the key parameters driving THz emission in this scenario are the plasma length $L$, over which THz emission takes place, and the dephasing length $l_d$, over which the two pump harmonics get out of phase. The latter length depends on the plasma density $N_e$. Other models describing the angular THz power spectrum have been elaborated in the literature; most of them also assume a homogeneous plasma cylinder \cite{KimNP,ThielePRA}. Although these assumptions are generally met, in practice, optical beam distortions are always present. They can affect the emitter's stability and symmetry, thereby degrading the generated THz radiation. An alternative theory \cite{Gorodetsky} tried to account for the role of plasma opacity, relevant for THz frequencies lower than the local plasma frequency, as well as oblique diffraction of THz waves through a transversally varying plasma channel and the experimental detector's spectral sensitivity. Although showing reasonable agreements between semi-analytical results and experimental measurements, this work does not include enough information about key parameters in THz emission (selected THz frequencies, plasma profiles, or the experimental detector frequency responses) to provide an accurate and convincing model for conical emission. In particular, this reference claims the emission is always conical at small THz frequencies. At the same time, one of the remaining problems to date is the seemingly generic coexistence of on-axis emission at low frequencies ($\lesssim 5$ THz) with a donut-shaped emission at large THz frequencies ($>5$ THz) in focused propagation \cite{Andreeva,Ushakov}.

This paper addresses this open question and is organized into three parts. Section \ref{sec:phys} introduces the underlying physics of THz generation, mainly dominated by photocurrents. It also recalls the so-called ''Local Current'' (LC) model, extended to include the instantaneous Kerr response. In section \ref{sec:model}, we establish a revised (''augmented'') THz emission model that keeps the key ingredients of the theory developed by You et al.~\cite{You2012} but accounts for (i) the spectral distributions characterizing the propagation regimes through three baseline configurations marked by different interplays between photocurrents and Kerr effect, (ii) transverse variations in the emitting source induced by laser spatial inhomogeneities or a simple linear focusing phase, and (iii) optical envelope profiles and changes in the laser polarization state, e.g., from linear to circular. Results obtained from this augmented model are compared with the THz spatio-spectral distributions provided by vectorial full space and time-resolved unidirectional pulse propagation (UPPE \cite{Kolesik2004}) simulations. Section \ref{sec:DM} is dedicated to the influence of optical aberrations on laser propagation, plasma geometry, and THz radiation. We model optical aberrations using Zernike polynomials that affect the input pulse phase \cite{Noll}. We then compare the resulting angular THz spectra evaluated by our semi-analytical model with UPPE simulation results for the most common optical aberrations in laser experiments: sphericity, astigmatism, and coma. Section \ref{sec:conclusion} concludes this work. 

\section{Physical mechanism for THz generation and laser propagation model}
\label{sec:phys}

We study the THz emission by two-color femtosecond (fs) light pulses that trigger air plasmas. The laser field can be either linearly polarized along the $x$-axis when the two harmonics are oriented along the same transverse direction (LP-P for "Linear Polarization - Parallel") or circularly polarized when the colors are circularly polarized and both rotate in the same direction (CP-S for "Circular Polarization - Same"). Previous work \cite{Meng,NJP} reported that CP-S laser pulses can increase by a factor $\sim 4-6$ the laser-to-THz conversion efficiency compared to their LP-P counterpart. The temporal profiles of the input pulse at $z=0$ are chosen as
\begin{align}
\label{ELPP}    \Vec{E}_{\rm LP-P} &= \left[\mathcal{E}_1 \cos(\omega_0 t) + \mathcal{E}_2 \cos(2 \omega_0 t + \phi)\right] \vec{e}_x \\
\label{ECPS}    \Vec{E}_{\rm CP-S} &= \frac{1}{\sqrt{2}} \binom{\mathcal{E}_1 \cos(\omega_0 t) + \mathcal{E}_2 \cos(2 \omega_0 t + \phi)}{\mathcal{E}_1 \sin(\omega_0 t) + \mathcal{E}_2 \sin(2 \omega_0 t + \phi)}
\end{align}
where $\omega_0$ is the laser fundamental frequency, $\phi$ the phase angle between the two colors and $\mathcal{E}_{1,2}$ their envelopes being initially Gaussian in time and space: 
\begin{equation}\label{Env}
    \mathcal{E}_{1, 2}(\vec{r}_\perp, t) = \sqrt{r_{1,2}} E_0 \exp \! \left[-\frac{t^2}{t_{1,2}^2}-\frac{r_\perp^2}{w_{1,2}^2}\right].
\end{equation}
Here $E_0$ is the laser amplitude, $t_{1,2}$ and $w_{1,2}$ are the $1/e^2$ duration and width of the two harmonics, and $k_{1,2}$ their corresponding wavenumbers. The peak intensity ratio of fundamental over second harmonic is determined by $r_2=r$ and $r_1 = \sqrt{1 - r}$, with $0 \le r \le 1$. A focusing lens with focal length $f$ is taken into account by adding a phase to the input pulse in the temporal Fourier domain,
\begin{equation}\label{lense}
\widehat{\vec{E}}(r_\perp,\omega) \quad \rightarrow \quad \widehat{\vec{E}}(r_\perp,\omega)\exp \! \left(- i \frac{\omega}{c} \frac{r_\perp^2}{2 f}\right).
\end{equation}
Later in Sect.~\ref{sec:DM}, the transverse spatial dependency of this phase term $\propto r_\perp^2/2 f$ describing a perfect focusing lens will be generalized to an $(x,y)$-dependent function $Z_{\rm OA}(\vec{r}_\perp)$, accounting for optical aberrations.

Photoionization caused by the laser pulse with intensity $\sim 10^{13-15}$ W/cm$^2$ takes place in the tunneling regime \cite{Keldysh, QST, PPT}, leading to a stepwise increase in the free electron density described by the rate equation
\begin{equation}\label{dtNe}
    \partial_t N_e = W(E) \left(N_a - N_e\right),
\end{equation}
where $W(E)$ is the photo-ionization rate, and $N_a$ is the initial neutral density (here written for one species for simplicity). The free electrons are then accelerated by the laser field and generate photocurrents governed for non-relativistic plasmas by
\begin{equation}\label{dtJe}
    \left(\partial_t + \nu_c \right) \Vec{J}_e = \frac{e^2}{m_e} N_e \Vec{E},
\end{equation}
where $\nu_c$ denotes an electron-neutral collision rate taken as $\nu_c = 2.86$~THz. Although low-frequency radiation is mainly generated from photocurrents, for weaker laser intensities ($< 10^{13}$ W/cm$^2$), an additional contribution to the overall THz yield may be provided by the Kerr polarization, that is, the third-order nonlinear response of the bound electrons \cite{Borodin2013,Andreeva,AliseeOE} reading
\begin{equation}\label{PKerr}
    \vec{P}_K(\vec{r}, t) = \varepsilon_0 \chi^{(3)} {E}^2(\vec{r}, t)\vec{E}(\vec{r}, t),
\end{equation}
with $\chi^{(3)}$ being the third-order electric susceptibility \cite{Boyd} assumed to be non-dispersive, i.e., $\chi^{(3)} \equiv \chi^{(3)}(-\omega_0,\omega_0,\omega_0)$. A second well-known action of the Kerr response is to transversally compress the laser beam and maintain it in a self-focused state through optical self-focusing and plasma defocusing sequences as its peak power remains higher than the critical power for self-focusing \cite{ROP2007},
\begin{equation}
    P_{\rm cr} \approx \frac{3.72 \lambda_0^2}{8 \pi n_0 n_2},
\end{equation}
with $n_0 = n_{\rm opt}(\omega_0)$, $n_{\rm opt}(\omega)$ denoting the linear optical refractive index of air taken from Ref.~\cite{Peck}, and $n_2=3\chi^{(3)}/(4n_0^2c\varepsilon_0)$ is the nonlinear refractive index.

During propagation in air, the laser pulse can also excite rotational transitions, mostly in $N_2$ molecules, leading to stimulated Raman rotational scattering (SRRS) with polarization \cite{EPJST}
\begin{equation}\label{PRaman}
\vec{P}_R(t) = \frac{3}{2} x_K \varepsilon_0 \chi^{(3)} \int_0^{+\infty} \vec{G}(\vec{E}, t, \tau) R(\tau) d\tau,  
\end{equation}
where $x_K$ is the ratio of SRRS to the total third-order response. $\vec{G}(\vec{E}, t, \tau)$ is a function describing the third-order interaction:
\begin{equation}
\vec{G}(\vec{E}, t, \tau) = E^2(t - \tau) \vec{E}(t) - \frac{2}{3} \binom{E^2_y(t - \tau)E_x(t)}{E^2_x(t - \tau)E_y(t)},
\end{equation}
and $R(\tau) = \tau_1 \left(\tau_1^{-2} + \tau_2^{-2} \right) \sin(\tau/\tau_1) e^{- \tau/\tau_2}$ is the delayed SRRS response with inverse rotational frequency $\tau_1 \approx 62.5$ fs, and $\tau_2 \approx 77$ fs refers to the characteristic dipole dephasing time \cite{Penano, Pitts}. The ratio $x_K$ is estimated to $\approx 80$~\% according to Zahedpour et al.~\cite{Zahedpour}. Because both the instantaneous Kerr response and the delayed SRRS response contribute to the third-order nonlinear polarization, a factor $(1 - x_K)$ must be applied in Eq.~(\ref{PKerr}) when accounting for SRRS. As shown in Ref.~\cite{EPJST}, the SRRS contribution to the overall THz yield is negligible compared to the instantaneous Kerr response and, above all, to the photocurrents.

In the following, we shall perform a systematic comparison between semi-analytically computed angular spectra and far-field THz spectra calculated using the unidirectional pulse propagation equation (UPPE) \cite{Kolesik2002, Kolesik2004},
\begin{multline} \label{UPPE}
    \frac{\partial \overline{\vec{E}}}{\partial z} = i \sqrt{k^2(\omega) - k_\perp^2}\ \overline{\vec{E}} \\+ \frac{i \mu_0 \omega^2}{2 k(\omega)}  \left[\frac{i}{\omega} \left(\overline{\vec{J}}_e + \overline{\vec{J}}_{\rm loss} \right) + \overline{\vec{P}}_{K} + \overline{\vec{P}}_{R} \right].
\end{multline}
Here, the notation $$\overline{f}(\vec{k}_\perp,\omega) = \iiint_{-\infty}^{+\infty} f(\vec{r}_\perp,t) e^{-i \vec{k}_\perp \cdot \Vec{r}_\perp + i \omega t} d\vec{r}_\perp dt$$
refers to the Fourier transform of any function $f(\vec{r}_\perp,t)$ in time and transverse variables, and $\vec{J}_{\rm loss}$ is the loss current density associated with photoionization. In our analysis and simulations, the propagation medium is ambient air, composed of 20 \% O$_2$ and 80 \% N$_2$ with ionization energies $U_i = 12.1$ eV \cite{NIST} and $U_i = 15.6$ eV \cite{NIST}, respectively. The initial neutral density is $N_a = 2.7 \times 10^{19}$ cm$^{-3}$ and photoionization is modeled using the ionization rate of Popov, Perelomov and Terent'ev (PPT) \cite{PPT} with effective charge numbers $Z_{O_2}^* = 0.53$ and $Z_{N_2}^* = 0.9$ according to Talebpour et al.~\cite{Talebpour}.

Two representative $1/e^2$ beam widths will be considered, namely $w_{1,2} = 250\,\mu$m for spatially narrow beams and $w_{1,2} = 2.5$ mm for broad beams. To simulate the latter in the long-range filamentation regime and save computational cost, a pre-processor was used, which integrates the linear part of Eq.~(\ref{UPPE}) before reaching the electric field threshold of $\sim 1$ GV/m ($\equiv 10^{11}$ W/cm$^2$) above which the (Kerr) nonlinearity kicks in. Because the pre-processor solves a linear propagation equation, the solution can be obtained analytically:
\begin{equation}
    \overline{\vec{E}}(\omega, z + \Delta z) = \overline{\vec{E}}(\omega, z) e^{i \Delta z \sqrt{k(\omega)^2 - k_\perp^2} }.
\end{equation}
Further Fourier interpolation was then carried out to refine the spatial resolution of the resulting data, which were used as initial conditions for the UPPE simulations to properly compute the subsequent nonlinear propagation. The spatial resolution used in the UPPE simulations was $\Delta x = \Delta y = 4\,\mu$m for narrow beams and $\Delta x = \Delta y = 5-7\,\mu$m for broad beams and up to $15\,\mu$m for the filamentation of beam profiles distorted by coma. The temporal resolution was $\Delta t = 100$~as, and a (co-moving) time window of $1.5$~ps guaranteed a spectral resolution of $\Delta \nu = 0.66$ THz.

Besides performing costly UPPE simulations, we also integrated the "Local Current" (LC) model \cite{KimOE, KimNP, BabushkinLC}, which neglects propagation effects and approximates the electric field emitted by the plasma and Kerr source terms as
\begin{equation}\label{LC}
    \widehat{\vec{E}}(\omega) \propto \left(- i \omega \widehat{\vec{J}}_e(\omega) - \omega^2 \widehat{\vec{P}}_K(\omega) \right),
\end{equation}
$\widehat{f}(\omega)$ denoting the Fourier transform of $f(t)$.

\section{THz emission model}
\label{sec:model}

Analytical results from the LC model provide the dependence on the phase angle $\phi$ of the THz amplitudes generated by photocurrents and Kerr response \cite{NJP} as
\begin{align}
    \label{ETHzLP} \vec{E}_{\rm THz, LP-P}^{\rm J} &\propto \sin \phi\ \vec{e_x}, & \vec{E}_{\rm THz, LP-P}^{\rm K} &\propto \cos \phi\ \vec{e_x}, \\
    \label{ETHzCP} \vec{E}_{\rm THz, CP-S}^{\rm J} &\propto \binom{\sin \phi}{\cos \phi}, & \vec{E}_{\rm THz, CP-S}^{\rm K} &\propto \binom{\cos \phi}{- \sin \phi}.
\end{align}
Here, $E_{\rm THz}^{\rm J}$ and $E_{\rm THz}^{\rm K}$ are the THz electric fields produced by the photocurrents (index J) and Kerr response (resp.\ K) of LP-P and CP-S pump pulses, respectively. In practice, plasma channels extend over millimeter to centimeter lengths in the focused regime and up to several meters in the (collimated or loosely focused) filamentation regime. Over sufficiently long propagation ranges, the optical and plasma dispersion induce an important variation of the phase $\phi$ along the plasma channel, i.e., for an axially homogeneous plasma:
\begin{equation}
\label{Phi_z}
    \phi(z) = \phi_0 + \pi \frac{z}{l_d},
\end{equation}
where $\phi_0$ is the phase angle taken at the plasma onset distance. The ''dephasing length'' $l_d$ represents the propagation distance over which $\phi$ changes by $\pi$ due to dispersion, that is,
\begin{equation}\label{eq:ld}
    l_d = \frac{\lambda_0}{4 \left[n(2 \omega_0) - n(\omega_0) \right]}
\end{equation}
with $\lambda_0 = 2 \pi c/\omega_0$. Here, $n(\omega)$ is the medium refractive index that accounts for the contributions of Kerr and plasma 
\begin{equation}\label{n_plasma}
    n^2(\omega) = n^2_{\rm opt}(\omega) + 2 n_0 n_2 I_0 - \frac{\omega_p^2}{\omega^2(1 + i \nu_c/\omega)},
\end{equation}
where $I_0 = c \varepsilon_0 n_0 E_0^2/2$ is the incident pulse intensity and $\omega_p = \sqrt{e^2 N_e/\varepsilon_0 m_e}$ is the electron plasma frequency. Here, we recall that $l_d$ depends on $N_e$ through the plasma frequency in Eq.~(\ref{n_plasma}). Note the difference by a factor of 2 with You et al.'s definition of the dephasing length, which, we believe, proceeds from a misprint in the original Ref.~\cite{You2012}. 

\subsection{You et al.'s conical emission model}

Before dwelling upon our revised model, we find it instructive to recall the main ingredients of You et al.'s \cite{You2012, You2013} theory that describes the THz intensity radiated in the far-field by a homogeneous ($N_e =$ const.) cylindrical plasma of length $L$ and radius $a$. Here, the only mechanism generating THz waves is the photocurrent density excited by an LP-P laser pulse, and thus, according to Eq.~(\ref{ETHzLP}), the local THz yield evolves as $\sin{\phi(z)}$. Derived from the radiated electric field in the far-field approximation, You et al.\ obtained the angular THz intensity spectrum (see Appendix for a detailed derivation):
\begin{multline} \label{You}
    |\widehat{E}(\omega_{\rm THz}, \Theta)|^2 \propto |A(\omega_{\rm THz})|^2 \left(\frac{J_1(\beta)}{\beta} \right)^2 \\ 
    \times L^2 \left[\kappa_+^2 + \kappa_-^2 - 2 \kappa_+ \kappa_- \cos(2 \phi_0) \right],
\end{multline}
where $A(\omega)$ is the spectral amplitude of the THz source, and $J_1$ is the Bessel function of first order. The quantities $\kappa_\pm$ and $\beta$ are given by
\begin{align}
\label{Kappa}\kappa_\pm &= \mathrm{sinc}\ \alpha_\pm \\
\label{Alpha} \alpha_\pm &= \frac{\pi L}{\lambda_{\rm THz}} \left(1 - \cos \Theta \pm \frac{\lambda_{\rm THz}}{2 l_d} \right)    \\
\label{Beta} \beta &= \frac{2\pi}{\lambda_{\rm THz}} a \sin \Theta, 
\end{align}
with $\mathrm{sinc}(x) = \sin x/x$ and $\lambda_{\rm THz}$ denotes the wavelength associated with the THz frequency $\omega_{\rm THz}$. The functions $\kappa_\pm$, linked to the phase matching condition for THz generation, depend on the extent $(L)$ of the axial filament. The quantity $\beta$ accounts for the radial dimensions of the plasma ($\beta \propto a$) entering the term $J_1^2(\beta)/\beta^2$ that describes the circular diffraction of the THz waves emitted by the plasma cylinder. For common filaments, the radius $a$ is limited to a few tens of $\mu$m and $\Theta < 10$°, i.e., $\beta \ll 1$ in the main part of the spectrum, so this diffraction term has limited influence. The maximum emission is reached for $\alpha_- = 0$, thus providing the theoretical emission angle $\Theta_p$:
\begin{equation}\label{Thetap_You}
    \cos \Theta_p = 1 - \frac{\lambda_{\rm THz}}{2 l_d}.
\end{equation}
The phase mismatch between the THz waves emitted at different places in the plasma filament causes destructive interference for the on-axis THz component for plasmas of length $L > l_d$, and the THz radiation propagates as a cone in space, far from the plasma volume. The emission angles are usually narrow, i.e., $\Theta$ is in the 2°~--~10° range~\cite{Andreeva, Rasmussen,Gorodetsky,Zhong2006}.

\begin{figure}
  \centering
  \includegraphics[width=0.9\linewidth]{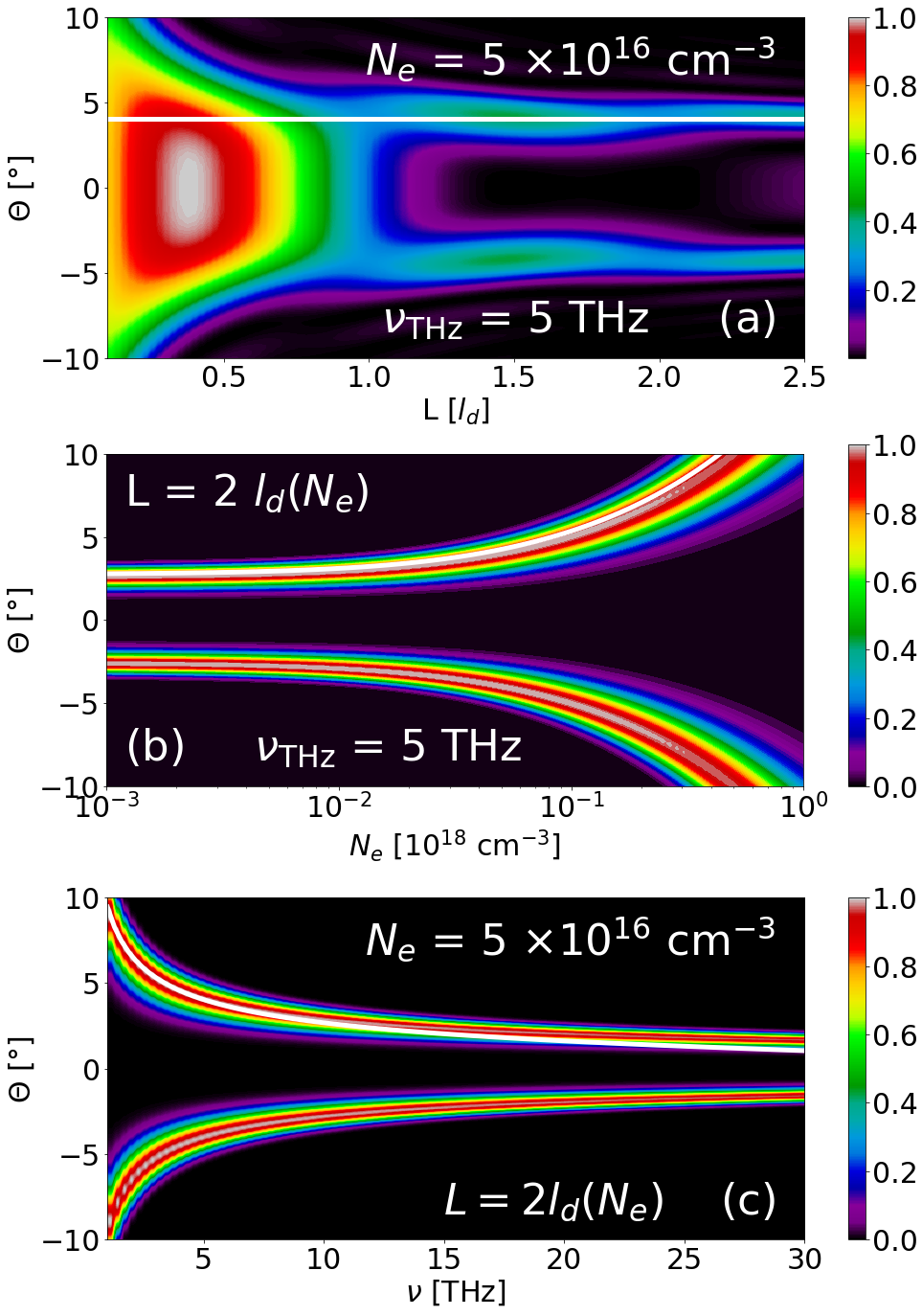}
\caption{\label{DependancesYou} (a,b,c) Mapping of the angular THz spectral intensity $|\widehat{E}(\omega_{\rm THz}, \Theta)|^2$ given by Eq.~(\ref{You}) as a function of the angle $\Theta$ and (a) the plasma length $L$, (b) the electron density  $N_e$, and (c) the frequency $\nu=\omega_{\rm THz}/2\pi$. The maps are normalized to their respective maximum values, and we set $A(\omega) = 1$ for simplicity (as in Ref.~\cite{You2012}). The white curves report the theoretical conical emission angle $\Theta_p$ predicted by Eq. (\ref{Thetap_You}).}
\end{figure}

As summarized in Fig.~\ref{DependancesYou}, the angular THz spectrum described by this model depends mainly on three parameters. The first is the length of the plasma $L$. When $L < l_d$, the THz radiation is maximal on the optical axis ($\Theta = 0$), while for longer plasma lengths ($L > l_d$), the on-axis THz amplitude vanishes and the THz beam becomes a cone with angle $\Theta_p$ [see Fig. \ref{DependancesYou}(a)]. The second key parameter is the electron density $N_e$. Figure~\ref{DependancesYou}(b) indicates that, at a fixed frequency, $\Theta_p$ increases for higher $N_e > 2 \times 10^{17}$ cm$^{-3}$ while it remains constant for lower density levels. The dephasing length $l_d$ [Eq.~(\ref{n_plasma})] decreases significantly as the value of $N_e$ increases. For larger density levels, conical emission, therefore, appears for shorter plasma lengths. The third parameter is the radiated THz frequency $\nu$. In this regard, Fig.~\ref{DependancesYou}(c) highlights that the emission angle $\Theta_p$ [Eq.~(\ref{Thetap_You})] strongly decreases with increasing $\nu$ (white curve), consistent with Eq.~(\ref{You}).

In addition to the above well-known properties, additional information can be extracted from the function describing the phase-matching condition in Eq.~(\ref{You}), namely, $P(L,\Theta,\lambda_{\rm THz}) \equiv L^2 [\kappa_+^2 + \kappa_-^2 - 2 \kappa_+ \kappa_- \cos(2 \phi_0)]$. First, on-axis components associated with the interference function $P(L,\Theta = 0,\lambda_{\rm THz}) = (2 l_d/\pi)^2 [1 - \cos{(\pi L/l_d)}] [1 - \cos{(2\phi_0)]}$ develop as a $2 l_d$-periodic function of the plasma length $L$ and reach maxima with constant amplitude at odd-numbered multiples of $l_d$, provided that the initial phase angle $\phi_0$ differs from 0 [$\pi$]. Second, phase-matched off-axis components parabolically increase like $P(L,\Theta =\Theta_p,\lambda_{\rm THz}) \simeq L^2 [1 + \mathrm{sinc}^2(\pi L/l_d) - 2 \cos(2 \phi_0)\mathrm{sinc}(\pi L/l_d)]$. They exceed the on-axis components from $L \approx l_d$ and increase monotonously to higher amplitude levels for longer plasma lengths. The value of $\phi_0$ does not change this behavior and simply introduces secondary oscillations. Importantly, these properties result from interference in the far-field THz intensity spectrum.

In the following, we revisit this THz emission model by making it as general as possible. This includes (i) treating envelope effects, (ii) allowing for transverse variations in the plasma density, (iii) accounting for the characteristic spectral signatures of $A(\omega)$ depending on the nonlinear propagation regime at play, and (iv) passing from LP-P to CP-S driver pulses.

\subsection{THz conical emission: An augmented model}

The electric field $\vec{E}(\vec{r}, t)$ emitted by a conductive current density $\vec{J}(\vec{r}, t)$ in a neutral plasma volume $V$ reads \cite{Jackson}
\begin{equation}
\label{29_2}
{\vec E}({\vec r},t) = - \frac{\mu_0}{4 \pi} \int_V d^3{\vec r}' \frac{\partial_t{\vec J}({\vec r}',t')}{|{\vec r} - {\vec r}' |},
\end{equation}
where $t' = t - |{\vec r} - {\vec r}' |/c$ is the retarded time at which the signal the detector receives at time $t$ was generated. We assume that the refractive index of the medium is $n_{\rm opt}(\omega_{\rm THz}) \equiv 1$ in the THz frequency range. The notation $\vec{r}'$ applies to the local coordinates inside the plasma volume. Cartesian, cylindrical, and spherical coordinates are written as ($x$, $y$, $z$), ($r$, $\varphi$, $z$) and ($r$, $\Theta$, $ \Phi$), respectively. We are interested in the far-field distribution, so the distance between the plasma and the detector is large compared to the dimensions of the filament, that is, $|\vec{r}| \gg |\vec{r'}|$, which leads to the simplification of the Fourier-transformed field (see Appendix~\ref{app1}):
\begin{multline}
\label{30bis_2}
\widehat{\vec E}(r,\Theta,\Phi,\omega) = \frac{i \omega \mu_0}{4 \pi r} \mbox{e}^{i \frac{\omega}{c}r} \int_{-L/2}^{+L/2} dz' \mbox{e}^{-i \frac{\omega}{c} \cos \Theta z'} \\ \times \iint d^2 \vec{r}'_\perp  \widehat{{\vec J}}({\vec r}',\omega)\mbox{e}^{- i {\vec k}_\perp \cdot {\vec r}'_\perp},
\end{multline}
with ${\vec k}_\perp = (\omega/c) (\sin\Theta \cos \Phi, \sin\Theta \sin \Phi)$. The double integral formally applies to the entire transverse plane. $\widehat{{\vec J}}({\vec r}',\omega)$ is zero for $|\vec{r'}|\rightarrow \infty$, in particular when considering a finite plasma cylinder of radius $a$.

The second step consists in taking the Kerr polarization [Eq. (\ref{PKerr})] into account through the substitution
\begin{equation}
\label{JK}
\vec{J}(\vec{r}, t) \rightarrow \vec{J}_e(\vec{r}, t) + \partial_t \vec{P}_K(\vec{r}, t),
\end{equation}
and calculating the free-electron contribution $\vec{J}_e(\vec{r}, t)$ using a toy ionization rate satisfying $W(E) \propto  E^2(t)$. This toy ionization rate was already employed in Refs.~\cite{Nguyen2018NJP,Tailliez2020} and allows a semi-analytical treatment that provides the major characteristics of multi-color-driven THz pulses, including their polarization state. Most importantly, it captures the asymmetries in the temporal location of the ionization peaks introduced by the second pump color, which are crucial for the photocurrent mechanism. In this respect, approaches based on Fourier series using more realistic, highly nonlinear ionization rates but involving only a single pump color \cite{Brunel1990,Kim2009} cannot account for these asymmetries.

The third step is to consider a pump field that admits transverse envelopes and different delays for the two pump colors. Introducing the retarded times $t_{j = \{1,2\}}' = t - z/v_{pj}$, with $v_{pj} = c/n(j \omega_0)$ being the respective phase velocities, our LP-P two-color laser field takes the general form
\begin{multline}
\label{35_2}
{\vec E}_{\rm L} ({\vec r},t) = {\vec {\cal{E}}}_1({\vec r}, t_1') \cos{[\omega_0 t_1' + \phi_1(\vec{r})]}\vec{e_x} \\ + {\vec {\cal{E}}}_2({\vec r}, t_2') \cos{[2 \omega_0 t_2' + \phi_2(\vec{r})]}\vec{e_x},
\end{multline}
where $\phi_j (\vec{r})$ is the spatial phase of the $j$th color. We assume that the envelopes $\mathcal{E}_j$ and their phases $\phi_j$ vary little along the filament, that is, they do not explicitly depend on $z$, and we can write $\mathcal{E}_j({\vec r}, t_1') \approx \mathcal{E}_j({\vec r}_\perp, t_1') $ and $\phi_j(\vec{r}) \approx \phi_j(\vec{r}_\perp)$. We note that this is a strong assumption, in particular for the phases, but is necessary for the upcoming semianalytical treatment. In contrast, our UPPE simulation results comprehensively account for the complete spatiotemporal dynamics of the pump field. Furthermore, we assume that temporal and radial variations of the laser envelopes are separable as
\begin{equation}
\label{40_2}
\mathcal{E}_j({\vec r}_\perp, t_j') = \mathcal{E}_{tr, j}({\vec r}_\perp) \mathcal{E}_{ax, j}(t_j'),
\end{equation}
where the maximum of $\mathcal{E}_{tr, j=1,2}$ is normalized to unity and $\mathcal{E}_{ax, j}$ contains the electric field amplitude expressed in V/m.

Straightforward computations detailed in Appendix~\ref{app1} then enable us to establish the following generic dependencies 
\begin{align}
\label{39_3}
\begin{split}
& \quad (\partial_t + \nu_c) \vec{J}_{e}^{\rm THz} \\
& \propto \mathcal{E}_1^2({\vec r}_\perp, t_1') \mathcal{E}_2({\vec r}_\perp, t_2')
\sin{\left[\Delta k z + \phi(\vec{r}_\perp) \right]} \vec{e_x},
\end{split}\\
{\vec P}_{K}^{\rm THz} & \propto \mathcal{E}_1^2 ({\vec r}_\perp, t_1')  \mathcal{E}_2({\vec r}_\perp, t_2')
 \cos{\left[\Delta k z + \phi(\vec{r}_\perp) \right]} \vec{e_x},
\end{align}
where $\Delta k$ is the phase mismatch in wavenumber between the harmonics and $\phi(\vec{r}_\perp)$ their phase angle:
\begin{align}
\label{39_2}
& \Delta k = \frac{2 \omega_0}{c}\left[ n(\omega_0) - n(2\omega_0) \right] = \pi/l_d,\\
& \phi(\vec{r}_\perp) \equiv \phi_2(\vec{r}_\perp) - 2 \phi_1(\vec{r}_\perp).
\end{align}
Further computations lead to
\begin{equation}
    \label{A-model}
    \left|\widehat{\vec E}({\vec r},\omega)\right|^2 = \frac{\mu_0^2}{16 \pi^2 r^2} \left|\vec{C}_J(\Theta, \Phi, \omega) + \vec{C}_K(\Theta, \Phi, \omega)\right|^2,
\end{equation}
where the functions $\vec{C}_J(\Theta, \Phi, \omega)$  and $\vec{C}_K(\Theta, \Phi, \omega)$ represent the contributions to the THz yield from the photocurrents (index $J$) and the Kerr effect (index $K$), respectively. These general functions are expressed as
\begin{equation}\label{eq:CJK}
    \vec{C}_{J, K}(\Theta, \Phi, \omega) = A_{J, K}(\omega) \vec{I}_{J, K}(\Theta, \Phi, \omega),
\end{equation}
$A_{J, K}(\omega)$ containing the frequency dependency of the considered source term and $\vec{I}_{J, K}(\Theta, \Phi, \omega)$ being the associated angle-dependent longitudinal phase matching function, namely,
\begin{align}
\label{IjSimp} \vec{I}_J^{\rm LP-P}(\Theta, \Phi, \omega) &= \frac{L}{2} \left[I_S^+ \kappa_+ - I_S^- \kappa_- \right] \vec{e_x}, \\
\label{INlSimp} \vec{I}_{K}^{\rm LP-P}(\Theta, \Phi, \omega) &= i \frac{L}{2} \left[I_S^+ \kappa_+ + I_S^- \kappa_- \right] \vec{e_x}.
\end{align}
Here, $\kappa_\pm$ are those given in Eq.~(\ref{Kappa}).

The quantities $I_S^\pm$ describe the transverse effects of the plasma on the far-field THz radiation, 
\begin{equation}\label{eq:ISpm}
I_S^\pm(\Theta, \Phi,\omega) = \mathcal{F}_\perp \left[J_{\rm tr}(\vec{r}_\perp') e^{\pm i \phi(\vec{r}_\perp')} \right](\Theta, \Phi,\omega),
\end{equation}
where notation $\mathcal{F}_\perp$ refers to the Fourier transform with respect to the transverse variables $\vec{r}_{\perp}'$ and 
\begin{equation}
\label{Jtr} J_{\rm tr}(\vec{r_\perp}) = \mathcal{E}_{\rm tr, 1}^2(\vec{r}_\perp) \mathcal{E}_{\rm tr, 2}(\vec{r}_\perp)
\end{equation}
represents the transverse envelope of the source terms. $I_S$ reduces to the Bessel-like circular diffraction function recalled in Eq. (\ref{You}) for a cylindrical homogeneous plasma of radius $a$ (see Appendix \ref{app2}). We also mention that CP-S pulses can be treated similarly, cf.\ Appendix \ref{app1}, and one finds
\begin{align}
\label{IjSimpCPS} \vec{I}_J^{\rm CP-S}(\Theta, \Phi, \omega) &= \frac{L}{2} \binom{I_S^+ \kappa_+ - I_S^- \kappa_-}{I_S^+ \kappa_+ + I_S^- \kappa_-}, \\
\label{INlSimpCPS} \vec{I}_{K}^{\rm CP-S}(\Theta, \Phi, \omega) &= i \frac{L}{2} \binom{I_S^+ \kappa_+ + I_S^- \kappa_-}{-I_S^+ \kappa_+ + I_S^- \kappa_-}.
\end{align}
By construction, our model does not account for variations in the plasma density along $z$, and plasma absorption and opacity are also neglected for reasons addressed in the appendix. 

\begin{figure*}
  \centering
  \includegraphics[width=.65\linewidth]{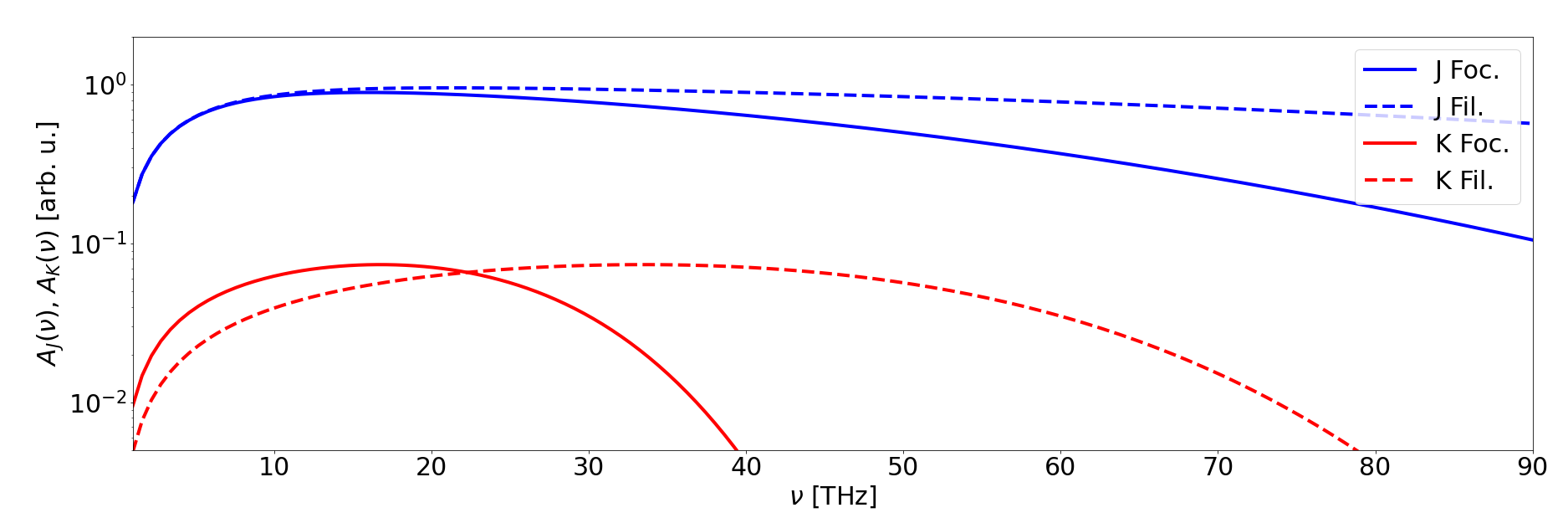}
\caption{\label{FigFormeA} Characteristic shapes of THz spectra due to photocurrents (blue curves) and Kerr effect (red curve) in focused (solid curves) and filamentation (dotted curves) regimes inspired from Refs.~\cite{AliseeOE} and \cite{Borodin2013}. The fraction of Kerr response can increase up to $10\%$ in amplitude in the filamentation regime or up to $77\%$ in the strong Kerr regime. Parameters in Eq.~(\ref{Aphot}) are $f_{\rm max} = 5$~THz and $f_0 = 60$~THz in focused regime and $f_{\rm max} = 5$~THz and $f_0 = 120$~THz in filamentation regime. Parameters in Eq.~(\ref{AKerr}) are $f_{\rm max} = 10$~THz and $f_0 = 30$~THz in focused regime and $f_{\rm max} = 5$~THz and $f_0 = 60$~THz in filamentation regime.}
\end{figure*}

Integration of our model requires some prior knowledge of key parameters of the emitting filament. First, we need to model the transverse shape of the THz source $J_{\rm tr}(\vec{r}_\perp')$ to calculate $I_S^\pm$. To this end, we use our linear laser propagation code up to a point $z = z_p$ where the width of the laser beam is small enough to reach high intensities corresponding to efficient plasma generation. At this distance, assuming that the envelopes of the two colors remain superimposed in time, we can extract the quantities $\phi({\vec r}_\perp)$ and the transverse field profiles, allowing us to reconstruct $J_{\rm tr}(\vec{r_\perp})$ at this distance and thus evaluate $I_S^\pm$. Second, as addressed in the appendix for illustrative purposes, Gaussian temporal envelopes impose a radiated Gaussian spectrum. This is unrealistic because the temporal pump pulse profiles strongly evolve along propagation due to the Kerr and plasma response and typical THz spectra obtained from experiments or simulations look very different. We therefore replace the profiles $A_{J, K}(\omega = 2 \pi \nu)$ in Eq.~(\ref{eq:CJK}) by typical spectral shapes delivered by the respective source term in the specific propagation regime under consideration. In what follows, we use
\begin{align}\label{Aphot}
    A_J(\nu) &\propto \left(1 - \mbox{e}^{-\nu/f_{\rm max}}\right) \mbox{e}^{-(\nu/f_0)^2} \\
    A_K(\nu) &\propto \left(1 - \mbox{e}^{-\nu/f_{\rm max}}\right) \mbox{e}^{-(\nu/f_0)^4} \label{AKerr}
\end{align}
yielding the synthetic profiles shown in Fig.~\ref{FigFormeA} for typical THz spectra produced by photocurrents alone (blue curve), or Kerr effect discarding SRRS (red curve) in either short-range focused (solid) and long-range filamentation (dashed curves) regimes. They have been inspired by Figs.~3(c,d) of Ref.~\cite{AliseeOE} and Fig.~2 of Ref.~\cite{Borodin2013} and are comparable to our transversely integrated UPPE spectra (not shown). Typically, plasma-driven THz spectra go up to $\sim 70 - 100$~THz, whereas bound electron responses only produce THz waves below $\sim 40$~THz in the focused regime. Their contribution may be extended to 80 THz in the long-range filamentation regime. In this self-guiding regime, nonlinearities develop a supercontinuum through which the THz spectrum can merge with the broadened optical pump spectrum~\cite{Berge2013}. Three typical interaction regimes will be explored in the following three subsections: purely plasma-driven emission in focused geometry, plasma regime interplaying with a strong Kerr response, and long-range filamentation regime.

\begin{figure*}
  \centering
  \includegraphics[width=0.65\linewidth]{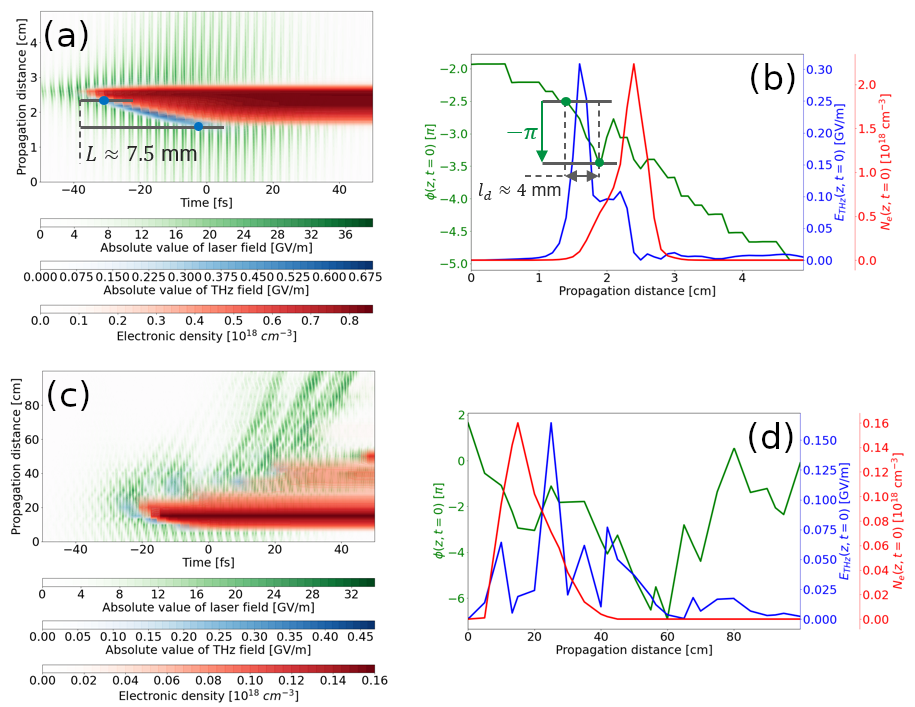}
\caption{\label{FigTHzMaps} Left-hand side column: on-axis THz fields (blue areas) generated by two-color laser fields (green curves) triggering electron plasma densities (red areas) in the $(z,t)$ plane in (a) focused plasma regime and (c) meter-range filamentation regime. Right-hand side column: (b,d) corresponding on-axis relative phase between the fundamental and second harmonic $\phi(z,t) = \phi_2(z,t) - 2 \phi_1(z,t)$ at $t = 0$ together with the created THz field and plasma density. The simulation parameters are detailed in the respective subsections below.} 
\end{figure*}

Lastly, the ''conical'' functions $\kappa_\pm$ are obtained by estimating the plasma length $L$ and the on-axis electron density $N_e$, which provides the dephasing length $l_d$. In estimating $N_e$ one has to remember that the THz waves are produced before the air plasma is fully established \cite{Rasmussen}, which means that the peak electron density (henceforth denoted by $N_e^{\rm max}$) is considerably higher than the effective electron density $N_e^{\rm eff}$ that should be used in the model. Illustrative examples obtained from UPPE simulations are given in Fig.~\ref{FigTHzMaps} detailing the maps of THz fields emerging from the interaction between the two-color laser field and the generated free electron density created in focused or extended propagation regimes. We can observe that in the focused regime [Figs.~\ref{FigTHzMaps}(a,b)] the THz field is always created at the front of the plasma, which justifies that the effective density level $N_e^{\rm eff}$ associated with the emitted THz wave is always lower than the peak plasma density. Another important observation is that the effective plasma length $L$ corresponding to the longitudinal distance along which the THz field is emitted remains approximately on the order of the distance along which the relative phase between the two laser harmonics undergoes a $\pi$ phase shift, that is, $L \gtrsim l_d$. By contrast, in the filamentation regime [Figs.~\ref{FigTHzMaps}(c,d)], we can see that although the THz pulse still emerges at the plasma front along Kerr self-focusing - plasma defocusing sequences, the plasma length generally satisfies $L \geq 2\,l_d$.

\subsection{Focused regime without Kerr response}

In this subsection, we compare the results of our augmented model with You et al.'s predictions and confront them with UPPE simulation results. We simulate the nonlinear propagation of a two-color LP-P Gaussian laser pulse with 0.8~µm fundamental wavelength, 500~$\mu$J energy, 34~fs duration, and 250~µm initial radius. 35\% of the laser energy is allocated to the second harmonic with zero relative phase. First, the pulse is focused in air by a lens of 2.6~cm focal length. The ionization is modeled with the PPT rate \cite{PPT} ($Z_{O_2}^* = 0.53, Z_{N_2}^* = 0.9$). To make photocurrents the only source of THz radiation in the simulation, we neglect the Kerr response and SRRS by setting the nonlinear index of air $n_2$ to zero.

\begin{figure}
    \centering
    \includegraphics[width=\linewidth]{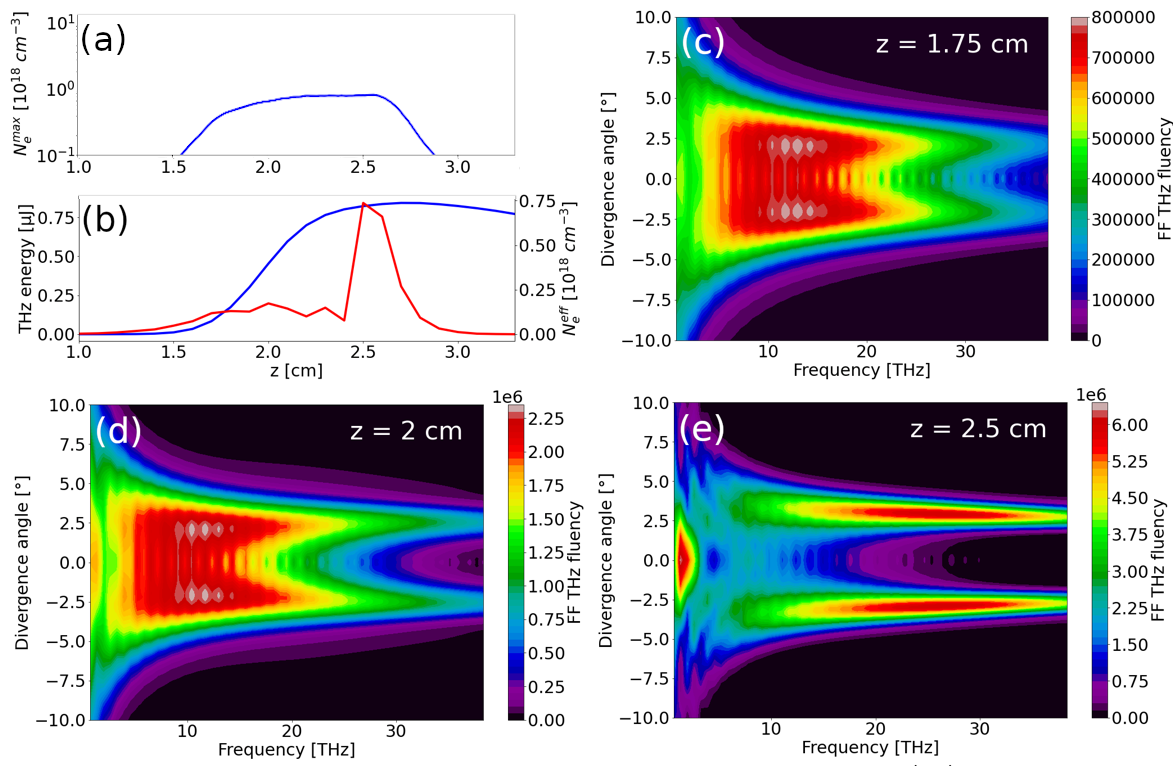}
    \caption{\label{CUP3DNoKerr} Results of UPPE simulations for laser-air parameters described in the text. (a,b) Evolutions along $z$ of (a) maximum electron density and (b) THz energy ($\nu < 90$~THz) in the simulation box (blue curve, left axis) and plasma density at the instants when the THz field is maximum (red curve, right axis). (c,d,e) Angular THz spectra in intensity calculated by transverse Fourier transform of the THz field at three different propagation distances.}
\end{figure}

Figures~\ref{CUP3DNoKerr}(a,b) plot the evolutions of the maximum electron density along $z$ and the THz energy contained in the simulation window, with the associated effective density $N_e^{\rm eff}$ at the time instants when the THz field is maximum. The peak density $N_e^{\rm max}$ is rather constant $\approx 10^{18}$~cm$^{-3}$ (resp. $N_e^{\rm eff} \approx 2 \times 10^{17}$ cm$^{-3}$) during the increase in THz energy, taking place between 1.5 and 2.25~cm. Figures~\ref{CUP3DNoKerr}(c,d,e) illustrate the angular THz intensity spectra calculated by Fourier transform in time and transverse space at $z = 1.75 $ cm (plasma onset), $z = 2$~cm (middle) and $z = 2.5$~cm (end of the emitting zone), respectively. Upon laser propagation, the initial on-axis THz radiation progressively transforms into a cone, starting from the upper part of the spectrum. This spectral pattern is often met in the literature \cite{Andreeva, Rasmussen}. The generic coexistence of on-axis emission at low frequencies with a donut-shaped emission at large THz frequencies in such focused propagation regimes \cite{Andreeva,Ushakov} is associated with an effective plasma length $L$ that remains close to the dephasing length $l_d$, as justified by Fig.~\ref{FigTHzMaps}(a,b). The spectrum plotted in Fig.~\ref{CUP3DNoKerr}(e) contains all the THz energy produced during the simulation. It presents a notable contribution on the optical axis for $\nu < 10-15$~THz, and conical emission at $\Theta \approx 3$° for $\nu > 15$~THz.

\begin{figure}
    \centering
    \includegraphics[width=\linewidth]{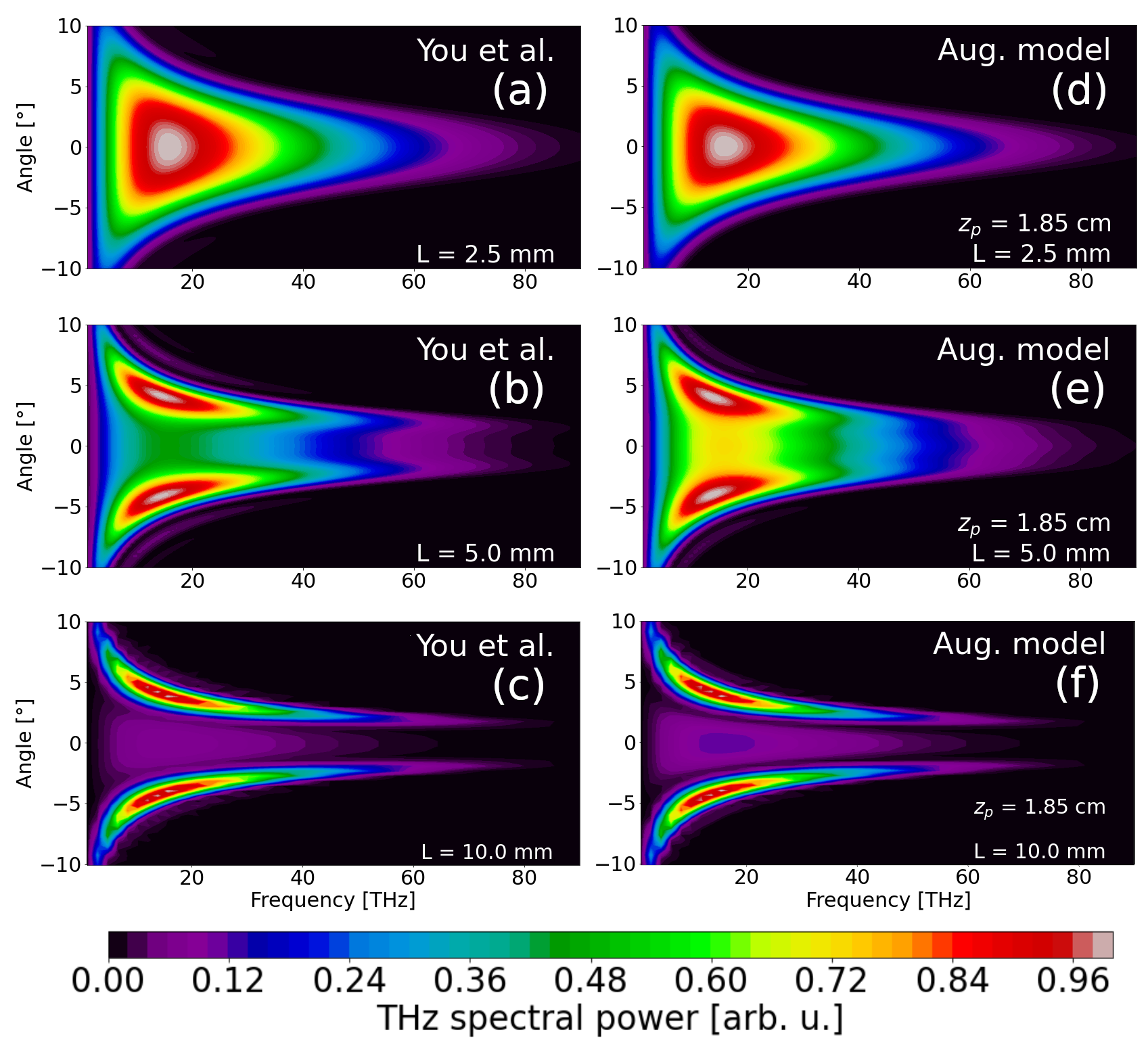}
    \caption{\label{FigYoudtJ} THz angular spectra evaluated by (a,b,c) You et \textit{al.}'s model and (d,e,f) our augmented model at (a,d) $z = 1.75$~cm $(L = 2.5$ mm), (b,e) $z = 2$~cm $(L = 5$ mm) and (c,f) $z = 2.5$~cm $(L = 10$ mm). The laser-gas parameters are specified in the text; $z_p = 1.85$ cm.}
\end{figure}

We now compare Figs.~\ref{CUP3DNoKerr}(c,d,e) to the angular intensity spectra predicted by You et \textit{al.}'s model and by our augmented model in Fig.~\ref{FigYoudtJ}. We assume that the only source of THz waves is photocurrents [$A_{\rm K}(\omega) = 0$]. Because we operate in the focused regime, the spectrum $A_J(\omega)$ is taken as the solid blue curve of Fig.~\ref{FigFormeA}. When integrating You et \textit{al.}'s model, the plasma radius is chosen as $a = 30$~µm and the effective electron density is $N_e^{\rm eff} = 2 \times 10^{17}$ cm$^{-3}$, yielding the dephasing length $l_d = 4$~mm. The maximum emitter length is determined by the growth of THz energy, thus $L = 0.75$~cm as justified above. When integrating our augmented model, we employ the same input laser-air parameters in our linear propagator as those used in the UPPE simulation. The relative phase $\phi_0 = \pi/4$ between the two colors is chosen at the input ($z=0$). We propagate the laser pulse linearly up to $z_p = 1.85$ cm ($f = 2.6$ cm), where the laser fluency takes a shape similar to that calculated by UPPE, i.e., it reaches a transverse Gaussian profile being 75~µm wide (not shown). The corresponding laser amplitude was checked to provide the same peak plasma density value ($N_e^{\rm max} \approx 10^{18}$ cm$^{-3}$) as in Fig.~\ref{CUP3DNoKerr}(a), and a 50~µm wide transverse plasma density profile.

Figure~\ref{FigYoudtJ} illustrates the results given by You et \textit{al.}'s model (a,b,c) and by our augmented model (d,e,f) at $z = 1.75$~cm, $2$~cm, and $2.5$~cm. At these distances, the effective plasma length $L$ is 2.5~mm, 5~mm and 10~mm, respectively. We notice that the spectra predicted by the two models in the cases where $L < l_d$ [Figs.~\ref{FigYoudtJ}(a,d)] and $L > l_d$ [Figs.~\ref{FigYoudtJ}(c,f)] are very similar, confirming that You and \textit{al.}'s model is well suited for radially symmetric plasma channels. However, for $L = 5~\textrm{mm} \approx l_d$, the strong on-axis contribution predicted by the UPPE simulation is better restored by the augmented model [Fig.~\ref{FigYoudtJ}(e) to be compared with Fig.~\ref{FigYoudtJ}(b)]. This difference comes from the phase modulation caused by the phase term $e^{\pm i \phi(\vec{r}_\perp')}$ in Eq.~(\ref{eq:ISpm}), which accounts for additional interferences between the THz waves formed in the overall plasma volume. This effect is absent in You et al.'s model. Note that for $L=10~\mathrm{mm} \approx 3\,l_d$, revivals of on-axis emission occur in this spectral dynamics -- in agreement with the interference function $P(L,0,\lambda_{\rm THz})$ discussed in the previous subsection -- and they are more pronounced in the augmented model.

\subsection{Focused regime with strong Kerr response}

We now examine what happens when a relevant Kerr contribution affects the former focused propagation. A second UPPE simulation was carried out with a strong instantaneous Kerr polarization by setting $x_K = 0$ and $n_2 = 3.8 \times 10^{-19}$ cm$^2$/W. This choice of $x_K$ implies that SRRS is ignored, which maximizes self-focusing and related four-wave mixing.

\begin{figure}
    \centering
    \includegraphics[width=\linewidth]{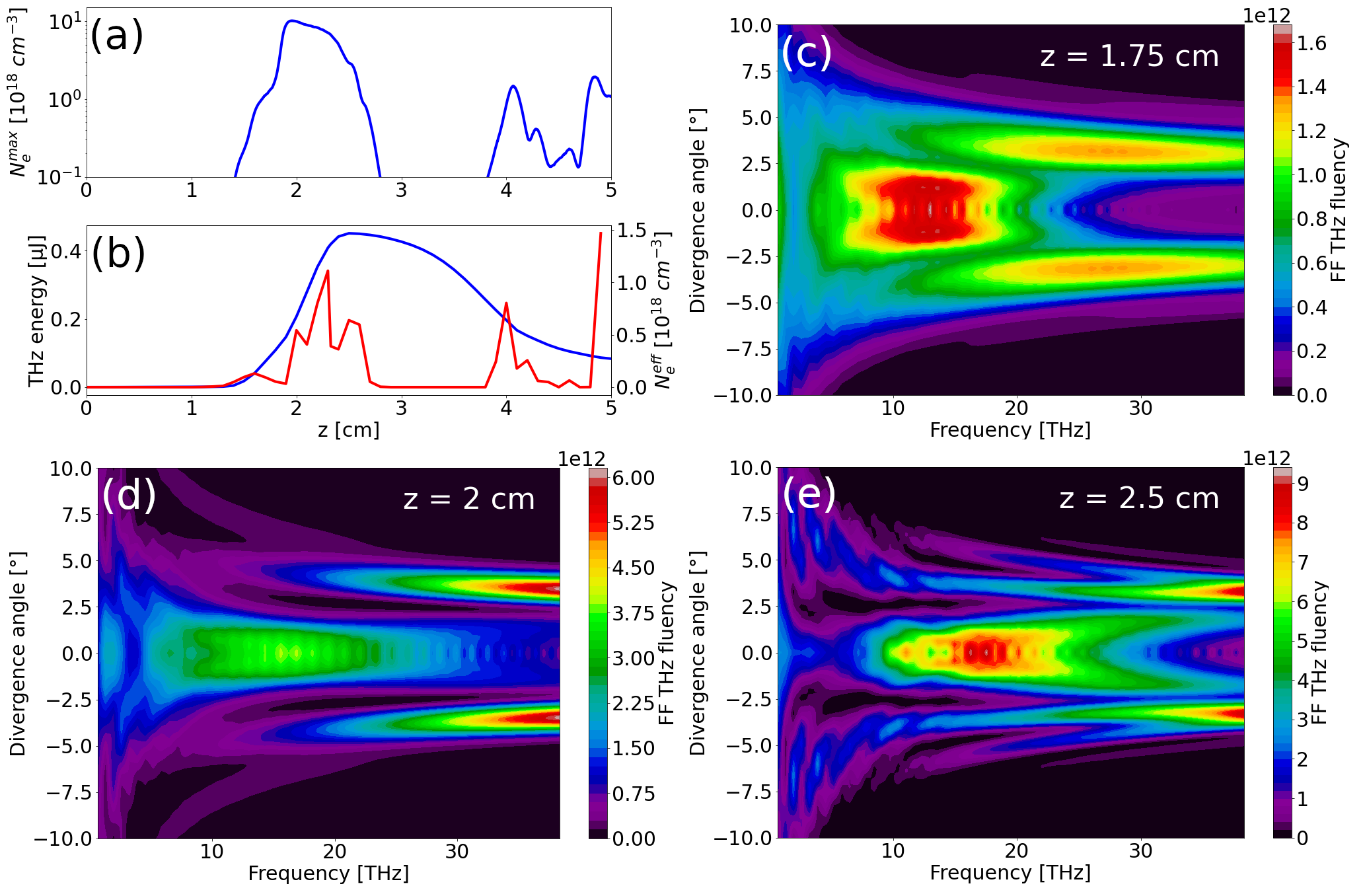}
    \caption{\label{CUP3DKerr} UPPE results displaying the same quantities as in Fig. \ref{CUP3DNoKerr} for a medium with strong Kerr response ($x_K = 0$, $n_2 = 3.8 \times 10^{-19}$ cm$^2$/W).}
\end{figure}

Figure \ref{CUP3DKerr}(a) shows that the maximum plasma density value, $N_e^{\rm max} \approx 10^{19}$ cm$^{-3}$, is about 10 times higher than in the previous configuration, which is caused by the strong Kerr self-focusing effect. Noticeably, this increase in density does not induce a growth in the laser-to-THz conversion efficiency [compare Figs. \ref{CUP3DNoKerr}(b) and \ref{CUP3DKerr}(b)]. The THz energy in the simulation box increases between 1.5 and 2.5~cm, hence $L = 10$~mm. Although the density strongly varies with $z$, we select $N_e^{\rm eff} \approx 3 \times 10^{17}$ cm$^{-3}$ [consistently with the red curve in Fig.~\ref{CUP3DKerr}(b)], which corresponds to a dephasing length $l_d = 2.8$~mm. Figures \ref{CUP3DKerr}(c,d,e) illustrate the angular THz spectra calculated by the UPPE code at $z = 1.75$~cm, $2$~cm, and $2.5$~cm. Again, the corresponding plasma lengths $L$ are $2.5$~mm, $5$~mm and $10$~mm. These spectra present strong on-axis components whose amplitudes increase along propagation. On-axis emission occurs in the frequency range $10-25$ THz, which we attribute to the strong Kerr self-focusing that affects the pump pulse propagation dynamics~\cite{AliseeOE}. Photocurrent-induced off-axis emission ($\Theta = 3-4$°) fully develops for high frequencies $\nu > 20$ THz in Figs.~\ref{CUP3DKerr}(d,e), where the criterion $L > l_d$ is fulfilled.

Now, let us evaluate the THz radiation using our semi-analytical model. The functions $A_J(\omega)$ and $A_{\rm K}(\omega)$ are taken as the solid blue and red curves of Fig.~\ref{FigFormeA}, respectively, with the Kerr maximum spectral amplitude being 77\% of the photocurrent one. Here, we linearly propagated the laser pulse up to $z_p = 2$~cm (instead of $z_p = 1.85$~cm) because stronger Kerr self-focusing fosters a more narrow plasma channel whose equivalent diameter is attained at longer distances $z$ upon linear propagation. The amplitude of the optical field reached at $z = z_p$ then corresponds to a peak electron density of $10^{19}$ cm$^{-3}$, in agreement with Fig.~\ref{CUP3DKerr}(a).

\begin{figure}
    \centering
    \includegraphics[width=\linewidth]{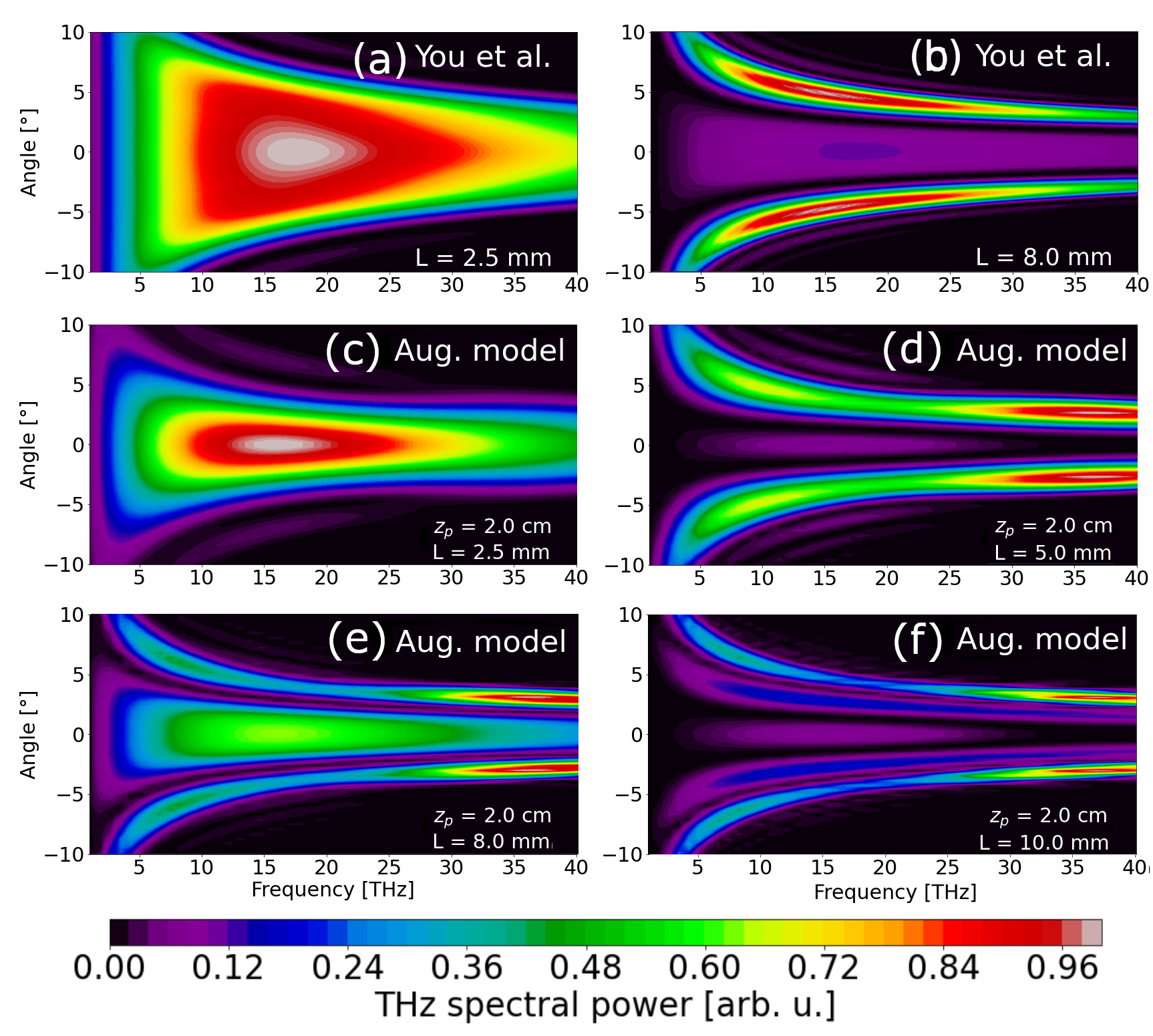}
    \caption{\label{YoudtJKerr} THz angular spectra evaluated by (a,b) You et \textit{al.}'s model and (c,d,e,f) our augmented model at (a,c) $z = 1.75$~cm $(L = 2.5$~mm), (d) $z = 2$~cm $(L = 5$~mm), (b,d) $z = 2.3$~cm $(L = 8$~mm), and (f) $z = 2.5$~cm $(L = 10$ mm); $z_p = 2$ cm. See text for details.}
\end{figure}

Figure~\ref{YoudtJKerr} shows the results of You et \textit{al.}'s model (a,b) and of the augmented model (c,d,e,f) for the same plasma lengths as analyzed in Fig.~\ref{CUP3DKerr}, completed with additional information for $L = 8$~mm. The two models predict similar conical emission angles. However, the angular spectra predicted by our augmented model include an extra contribution on the optical axis, peaked around $\nu = 15$~THz, which is shaped by the spectral amplitude $|A_J(\omega) - i \omega A_K(\omega)|$ where the parabolic Kerr contribution, here artificially increased, is absent in You et \textit{al.}'s model. In the full plasma regime ($z \approx 2.3$~cm), Figs.~\ref{YoudtJKerr}(b,e) both describe a net conical emission due to photocurrents at angles between 3 and 5°. Yet, only the augmented model produces off-axis emissions ($\Theta \lesssim 3.5$°) that develop in the high-frequency range $> 25$~THz, similarly to the UPPE simulations. On-axis revivals are more pronounced for $L = 8$~mm~$\simeq 3 \,l_d$ in the augmented model and reach a maximum at $\nu \approx 15$~THz, due to our parabolic spectral profile. Radial phase variations $\phi(\vec{r}_\perp)$ are also expected to contribute on-axis, following Figs.~\ref{FigYoudtJ}(e,f) of the previous subsection. The difference with UPPE data may be explained by the strong nonlinearities producing Kerr self-modulation and plasma defocusing, which render the plasma density in the UPPE simulation inhomogeneous, hence contradicting the model assumptions. On-axis components decrease to lower amplitudes in the semi-analytical spectra whereas they increase along $z$ in the UPPE simulation data. Finally, we underline that the angular spectra computed from the augmented model become closer to You et al.'s patterns when the Kerr spectral amplitude decreases below $50\%$ of the photocurrent spectral amplitude. This indicates that such large on-axis components developed in the UPPE simulations cannot be caused directly by the Kerr response alone.

\subsection{Filamentation regime}

We conclude this section by addressing the long-range filamentation regime. Here, the UPPE code was used to simulate a two-color laser pulse ($r = 35$\%) with fundamental operating at 0.8~$\mu$m, duration of 34~fs, initial radius of 2.5~mm, and energy of 1.5~mJ, loosely focused in the air (20\%~O$_2$, 80\%~N$_2$ - $n_2 = 3.8 \times 10^{-19}$~cm$^2$/W, $x_K = 0.8$) by a lens of focal length $f = 2.6$ m. For such a beam, using the linear pre-processor was necessary to perform a linear propagation of the laser pulse over 2.3~m until reaching intensities close to the Kerr nonlinearity threshold. The resulting pulse was then used as the input condition for the UPPE code at $z = 0$.

\begin{figure}
    \centering
    \includegraphics[width=\linewidth]{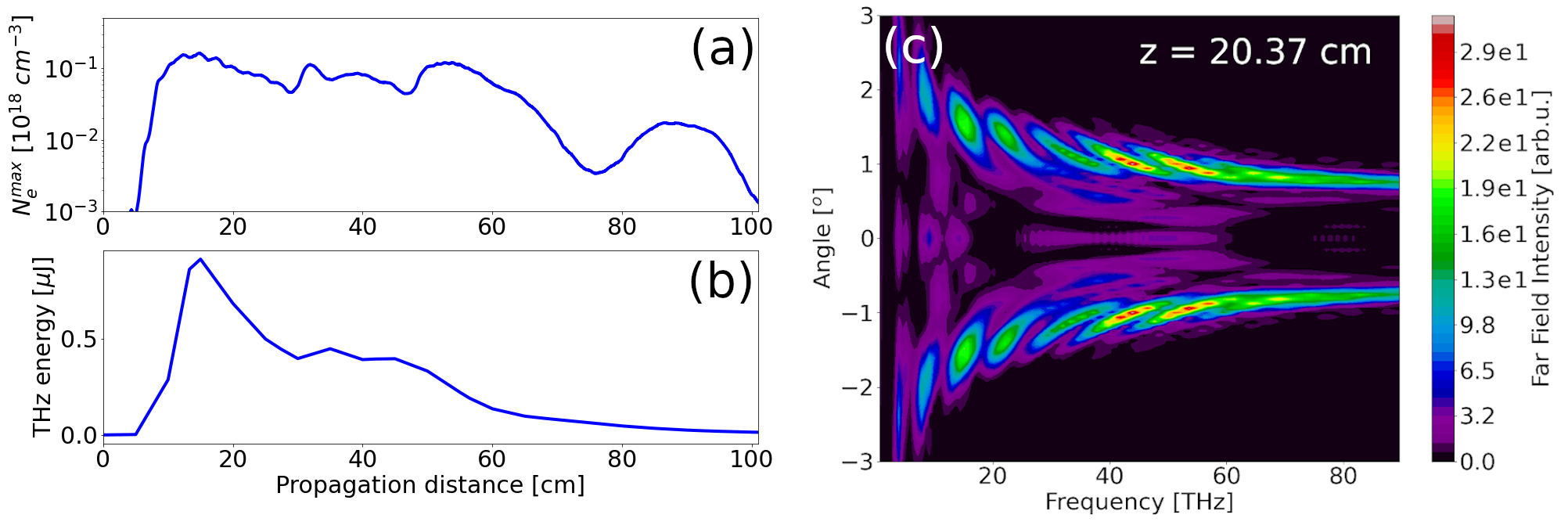}
    \caption{\label{CUP3DFil} (a) Maximum electron density and (b) THz energy ($\nu < 90$ THz) contained in the UPPE simulation window as a function of $z$ for a two-color long-range filament. (c) Angular THz spectrum calculated at $z \simeq 20.4$~cm.}
\end{figure}

 No multiple filamentation occurs (not shown) as the input peak power remains rather close to critical $P_{\rm cr}$ \cite{Akturk}, i.e., $P_{\rm in} \simeq 2.6\,P_{\rm cr}^{\rm inst}$ where $P_{cr}^{\rm inst} \equiv P_{cr}/(1-x_k)$ refers to the instantaneous critical power. Figure~\ref{CUP3DFil}(a) shows that the extended filament maintains a relatively constant maximum electron density $\approx 10^{17}$~cm$^{-3}$ between $z = 10$~cm and $z = 60$~cm, due to clamping effect, then turns off afterward. Figure~\ref{CUP3DFil}(b), plotting the THz energy in the simulation box along $z$, shows a strong decrease beyond $z = 18$~cm, due to the THz radiation leaving the simulation box. Moreover, THz production becomes less efficient at such long distances because the temporal overlap between the two laser harmonics deteriorates through temporal walk-off. Figure~\ref{CUP3DFil}(c) describes the THz spectrum calculated at $z\simeq 20.4$~cm. We can evaluate from Fig.~\ref{FigTHzMaps}(c) and Figs.~\ref{CUP3DFil}(a) and \ref{CUP3DFil}(b) a minimum length $L_{\rm min} \simeq 5$~cm associated with an efficient plasma emitting zone in this configuration. This length corresponds to the distance that a THz wave generated on-axis ($x = y = 0$) at an angle $\Theta \approx 1.5$° must travel to leave the simulation box with a transverse window extending from $-1.3$~mm to $1.3$~mm.

\begin{figure}
    \centering
   \includegraphics[width=\linewidth]{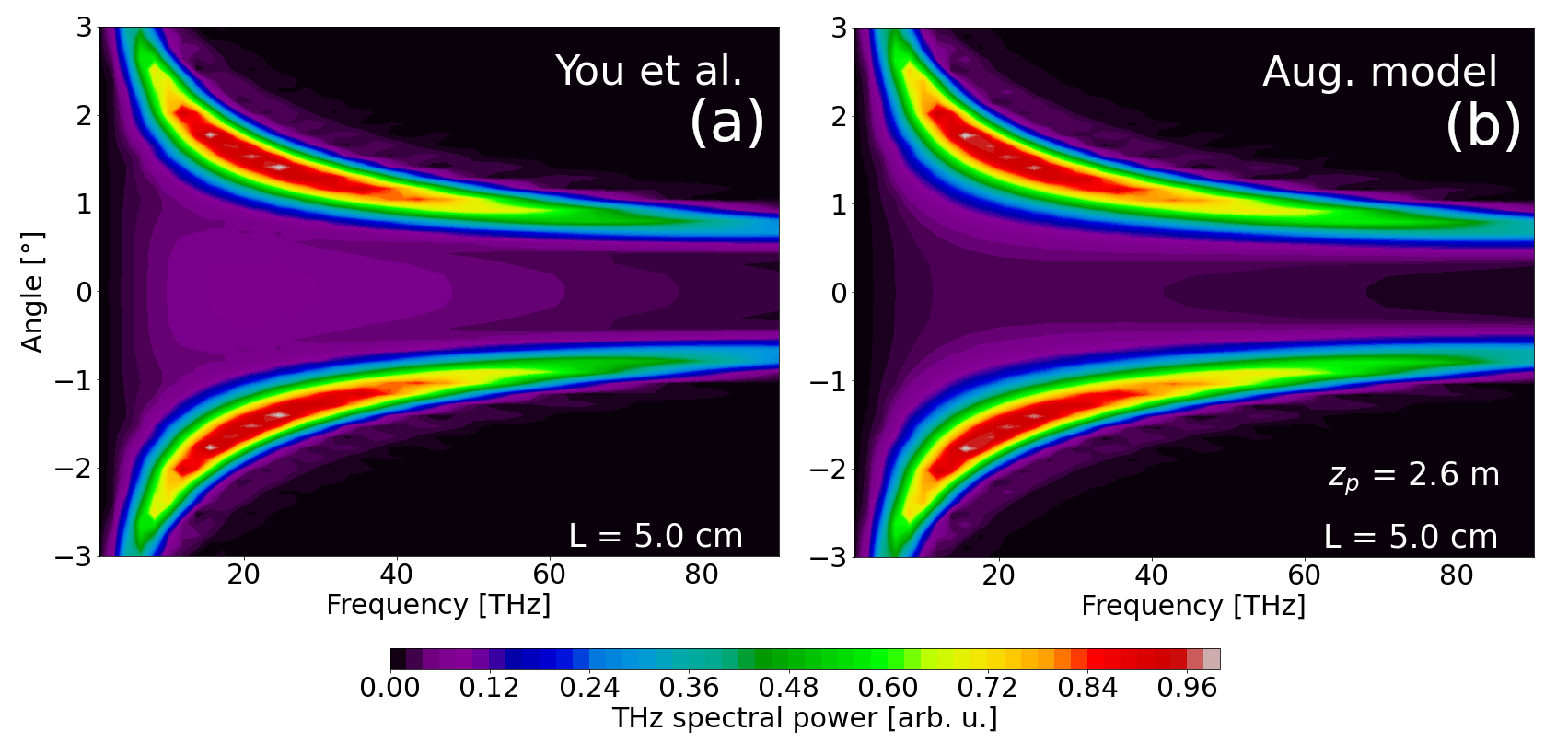}
    \caption{\label{dtJFil} THz angular spectra estimated using (a) You et \textit{al.}'s model and (b) our augmented model for the filamentation setup. Laser-medium parameters are specified in the text.}
\end{figure}

Setting $L = 5$~cm, the effective electron density can be estimated as $N_e^{\rm eff} \approx 10^{16}$~cm$^{-3}$ from Figs.~\ref{CUP3DFil}(a,b). Note that for such density levels, the dephasing length depends little on $N_e^{\rm eff}$ so that $l_d = 2.6$~cm~$< L$. We linearly propagated the same laser pulse over a distance of $z_p = 2.6$~m to integrate our augmented model. We chose $A(\omega)$ and $A_{J,K}(\omega)$ as given by Fig.~\ref{FigFormeA} (dashed curves), where the Kerr component is about $10\%$ that of photocurrents in intensity. Figures~\ref{dtJFil}(a,b) plot the angular spectra calculated from You et al.'s model and ours, respectively. These spectra are almost identical: They present a conical emission for $\Theta = 1-2$° with a flat frequency profile, which agrees with the UPPE simulation. The residual central emission $(< 10\% $ in spectral power) is consistent with the expected on-axis contribution for $L < 2\,l_d$.

\section{Influence of optical aberrations}
\label{sec:DM}

This section aims to characterize the impact of common optical aberrations (OAs) on the nonlinear propagation of two-color pulses producing THz radiation. OAs modify the propagation of a laser beam through distortion of its transverse phase, thus of its optical wavefront \cite{Born}. They tend to alter the position and shape of the focal spot and, thereby, the geometry of the plasma channel. OAs are usually classified using two integers: the radial degree $n$ and the azimuthal degree $m \leq n$ entering the Zernike polynomial $Z_n^m(\rho, \Phi)$ \cite{Noll}:
\begin{equation}
Z_n^m(\rho, \Phi) = \left\{\begin{array}{ll}
R_n^m(\rho) \cos(m \Phi) & \mathrm{if\ } m \geq 0 \\
-R_n^{-m}(\rho) \sin(m \Phi) & \mathrm{if\ } m < 0
\end{array} \right. ,
\end{equation}
where $\rho \equiv \sqrt{2} r_\perp/w_0$ with $r_\perp = \sqrt{x^2+y^2}$ and $w_0$ being the beam width.
\begin{equation}
R_n^m(\rho) = \sum_{k = 0}^{\frac{n - m}{2}} \frac{(-1)^k (n - k)!}{k! \left(\frac{n + m}{2} - k \right)! \left(\frac{n - m}{2} - k \right)!} \rho^{n - 2 k}
\end{equation}
is the radial distribution for an even number of $n-m$, and $R_n^m(\rho) \equiv 0$ otherwise.

Below we shall concentrate on the basic configuration of focus ($n = 2$, $m = 0$) - consisting of a simple focusing lens effect - and will add to this elementary lensing three characteristic OAs of interest, namely, vertical astigmatism ($n = m = 2$), vertical coma aberration ($n = 3$, $ m = -1$) and sphericity ($n = 4$, $n = 0$). For these four configurations, the polynomials $Z_n^m(\rho, \Phi)$ express as:
\begin{align}
\label{RFoc}&\textbf{Focus\:} & Z_2^0(\rho) = 2 \rho^2 - 1, \\
\label{RAstig}&\textbf{Astigmatism\:} & Z_2^2(\rho, \Phi) = \rho^2 \cos(2 \Phi), \\
\label{RComa}&\textbf{Coma\:} & Z_3^{-1}(\rho, \Phi) = (3 \rho^3 - 2 \rho) \sin(\Phi), \\
\label{RSph}&\textbf{Sphericity\:} & Z_4^0(\rho) = 6 \rho^4 - 6 \rho^2 + 1. 
\end{align}

Each OA is associated with an amplitude coefficient, $c_n^m$, which has the dimension of a distance and determines the impact of the respective OA on the transverse phase of the beam. The total phase distortion is then conveniently expressed as
\begin{equation} \label{Phia}
 -\frac{\omega}{c} Z_{\rm OA}(r_\perp,\Phi) = -\frac{\omega}{c} \sum_{n, m \leq n} c_n^m Z_n^m \left(\frac{\sqrt{2} r_\perp}{w_0}, \Phi \right).
\end{equation}
The polynomial $Z_{\rm OA}(r_\perp, \Phi)$ henceforth substitutes the transverse spatial dependency of the phase term $r_\perp^2/2f$ in Eq.~(\ref{lense}). The sum symbol means that when considering OAs such as sphericity, astigmatism, or coma, we add the latter to the baseline focus configuration.

\begin{figure}
	\centering
	\includegraphics[width=\linewidth]{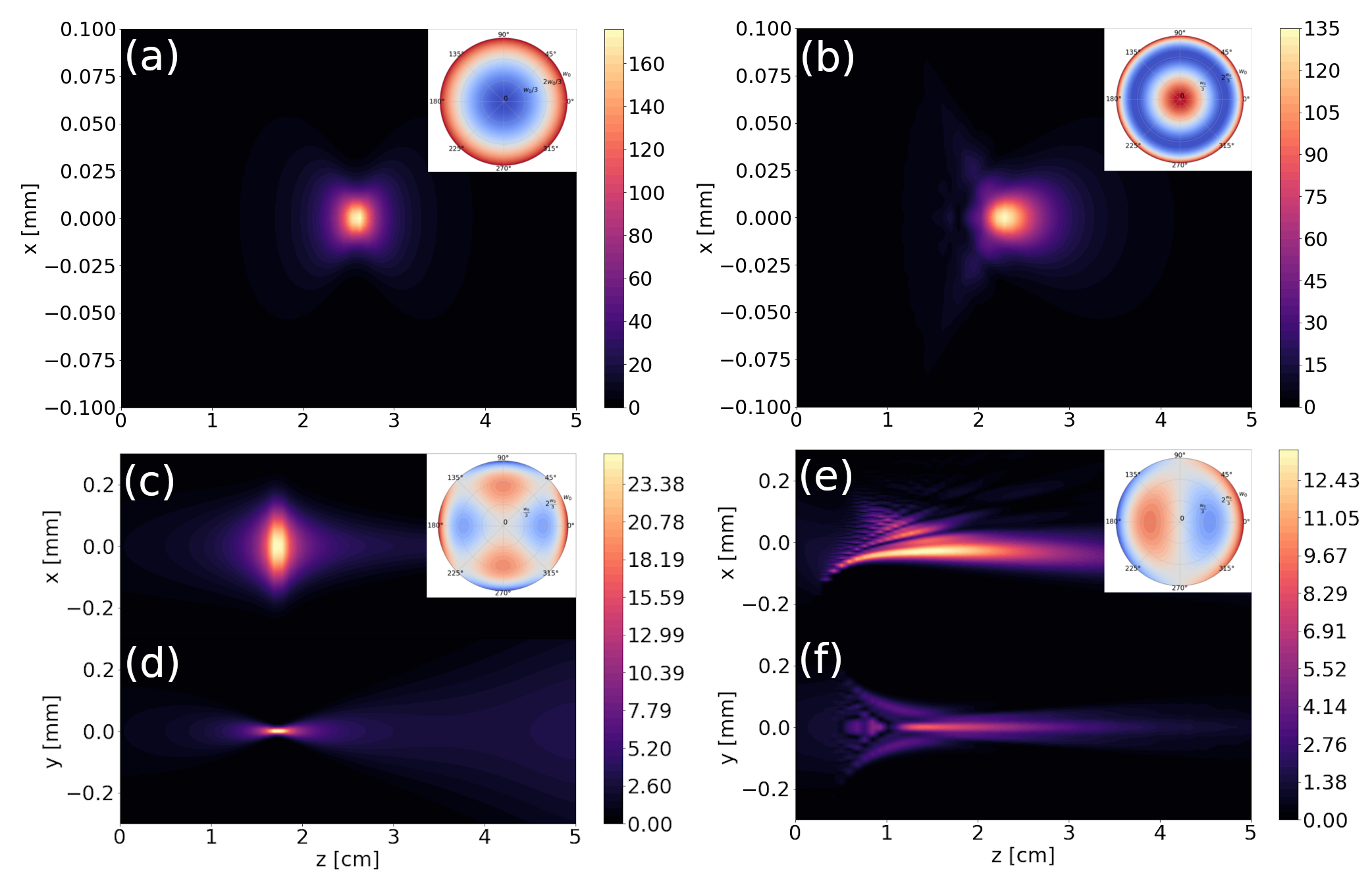}
	\caption{\label{PropLinDM} Laser fluency maps in the linear regime (a,b,c,e) in the ($x$, $z$) plane and (d,f) in the ($y$, $z$) plane for a two-color laser pulse with 800-nm fundamental ($r = 35$ \%) and initial radius $w_0 = 250$~µm focused in the air by a lens of focal length $f = 2.6$~cm (a) without additional optical aberration and with added (b) sphericity ($c_4^0 = 5$~nm), (c,d) astigmatism ($c_2^2 = 300$~nm) and (e,f) coma ($c_3^{-1} = 300$~nm). Insets show the corresponding laser spatial phases at $z = 0$.}
\end{figure}

Figure \ref{PropLinDM}(a) depicts linear laser propagation in the absence of OA (i.e., with focus only). The initially 250-µm wide optical pulse is focused in air by a lens with focal length $f = 2.6$~cm. The inset shows the initial spatial phase of the laser pulse, corresponding to the Zernike polynomial $Z_2^0(r_\perp)$ linked to the focusing effect. This reference case is compared with sphericity ($c_4^0 = 5$~nm), astigmatism ($c_2^2 = 300$~nm) and coma ($c_3^{-1} = 300$~nm), further added to focus.

Figures \ref{PropLinDM}(b-f) evidence that laser propagation is strongly influenced by the OAs. In particular, astigmatism and coma introduce transverse asymmetries. In the following, we discuss nonlinear UPPE simulation results characterizing the impact of astigmatism, coma, and sphericity in the focused regime. We begin by addressing LP-P laser fields, which we will compare to circularly polarized beams (CP-S) simulations. Finally, we present results in the filamentation regime for OA-influenced LP-P laser pulses. UPPE far-field spectra are compared with the results of our augmented model.

\subsection{LP-P Polarization}

Our two-color laser driver again involves a $0.8$~µm fundamental pulse with 34-fs duration and 250 µm initial radius. The laser energy is 500 µJ, with $r=35\%$ in the second harmonic; the pump beam is linearly polarized (LP-P) along $\vec{e_x}$ and focused ($f = 2.6$ cm equivalent to $c_2^0 = 300$ nm) in air ($n_2 = 3.8 \times 10^{-19}$ cm$^2 $/W, $x_k = 0.8$) to provide us with the reference case (focus). We then carried out three other simulations including sphericity ($c_4^0 = 5$ nm), astigmatism ($c_2^2 = 300$ nm) and coma ($c_3^{-1} = 300$ nm) imposed at $z = 0$. 

\begin{figure}
	\centering
	\includegraphics[width=\linewidth]{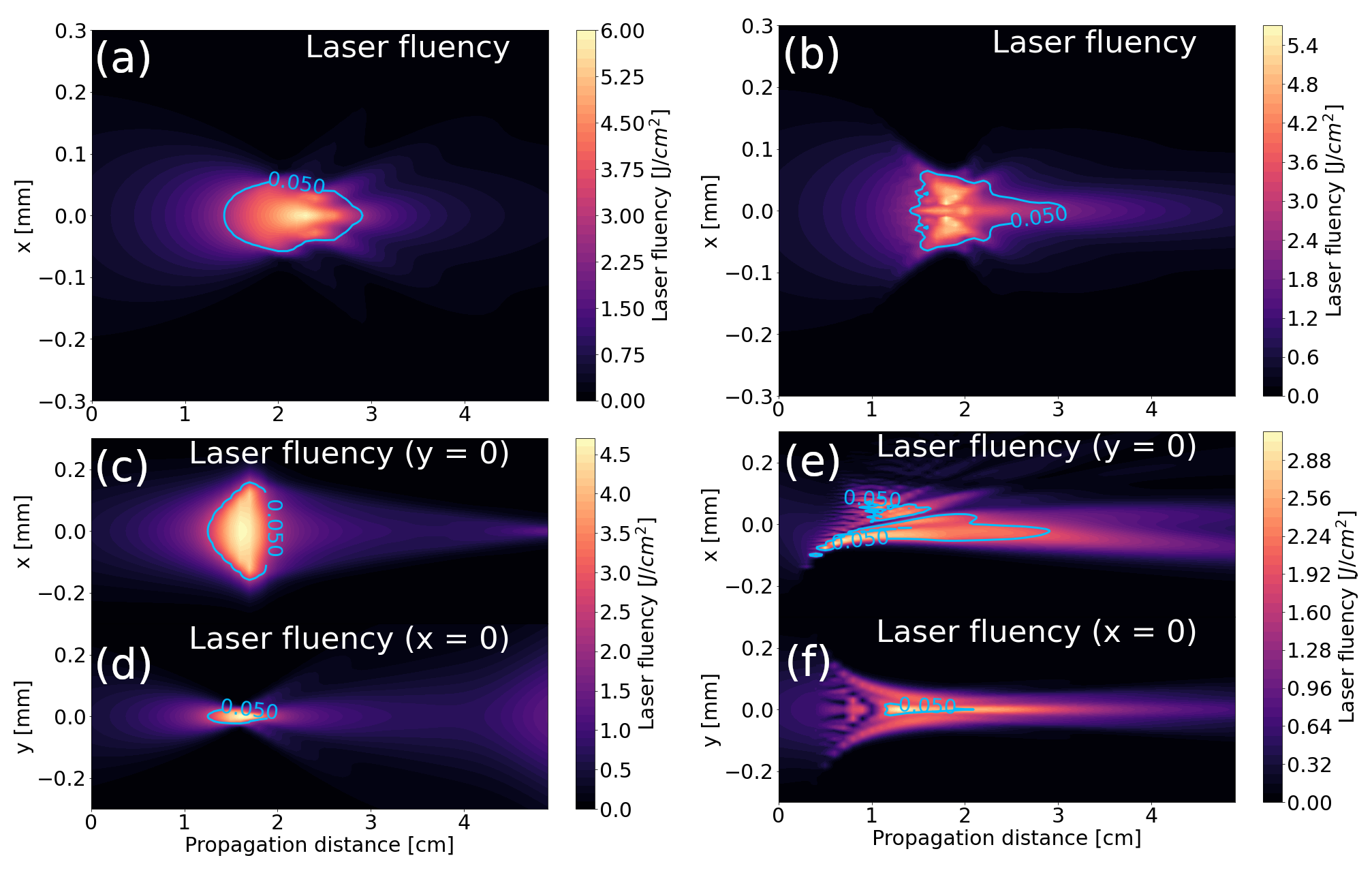}
	\caption{\label{E_NeLPP} UPPE-computed maps of the laser fluency for a two-color LP-P pulse propagating in nonlinear regime (a,b,c,e) in the ($x$, $z$) plane and (d,f) ($y$, $z$) plane (a) in the reference (focus) case, with added (b) sphericity, (c,d) astigmatism and (e,f) coma aberration. The blue lines show density iso-contours at $N_e^{\rm max} = 5 \times 10^{16}$ cm$^{-3}$. Simulation parameters are specified in the text.}
\end{figure}

To start with, Fig.~\ref{E_NeLPP} shows the evolution of the laser fluencies calculated with the UPPE code in the nonlinear regime. The propagation dynamics look similar to those observed in the linear regime (compare with Fig.~\ref{PropLinDM}). Differences (reduction) in the maximum fluency values can be attributed to the clamping of the maximum laser intensity by the peak plasma density, as indicated by the blue iso-contour density levels $N_e^{\rm max} = 5 \times 10^{16}$ cm$^{-3}$. For the reference case [Fig. \ref{E_NeLPP}(a)] and in the presence of sphericity [Fig. \ref{E_NeLPP}(b)], the plasma remains axisymmetric. However, adding astigmatism [Figs.~\ref{E_NeLPP}(c,d)] and coma [Figs.~\ref{E_NeLPP}(e,f)] breaks the transverse symmetry. Note that in the configuration with coma, the plasma is, in fact, composed of a filament extending along the laser propagation direction, accompanied by small off-axis zones with non-zero electron density around $z = 1$ cm.

\begin{figure}
	\centering
	\includegraphics[width=\linewidth]{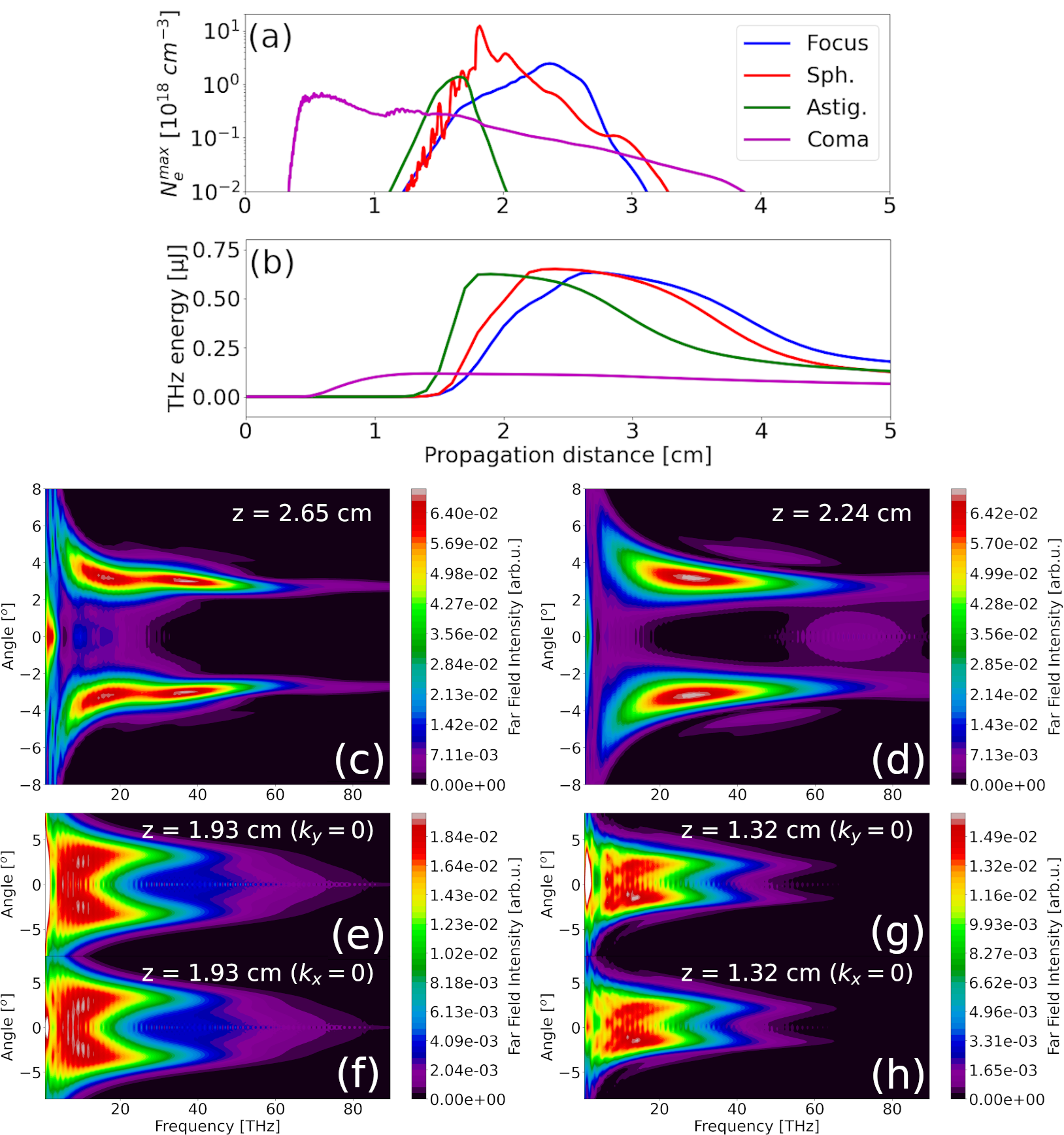}
	\caption{\label{ConesFocLPP} Top: (a) Maximum electron density and (b) THz energy ($\nu < 90$ THz) along $z$ in the reference (focus) case (blue curve), with sphericity (red curve), astigmatism (green curve) and coma aberration (magenta curve). Bottom: Angular THz spectra calculated with the UPPE code (c,d,e,g) in the ($x$, $z$) plane and (f,h) in the ($y$, $z$) plane (c) in the focus case at $z = 2.65$~cm, (d) in the sphericity case at $z = 2.24$~cm, (e,f) for astigmatism at $z = 1.93$~cm and (g,h) for coma at $z = 1.32$~cm.}
\end{figure}

Figure~\ref{ConesFocLPP}(a) describes the axial evolutions of the maximum electron density for the four considered configurations. Compared to focus (reference case, blue curve), reaching a peak electron density of $N_e^{\rm max} \approx 2-3 \times 10^{18}$~cm$^{-3}$, astigmatism (green curve) and coma aberration (magenta curve) reach lower density levels by a factor of $\approx 2$ and $\approx 5$, respectively. Sphericity generates very intense density peaks ($N_e \approx 10^{19}$ cm$^{-3}$) corresponding to the off-axis hot spots visible in the laser fluence map of Fig.~\ref{E_NeLPP}(b) at $z \simeq 1.8-2$~m. Sphericity produces a longer plasma channel than is observed in the focus case, which is in accordance with the references \cite{Ionin, Apeksimov}. Astigmatism leads to a shorter plasma because this aberration induces an additional focusing effect. Coma generates a longer and smoother plasma channel because it decreases the maximum fluency level (and, therefore, the peak electron density). Plasma generation starts at smaller $z$ distances, in accordance with Fig.~\ref{PropLinDM}(e,f). Figure~\ref{ConesFocLPP}(b), plotting the THz energy ($\nu < 90$ THz) accumulated along $z$ in the simulation window, shows that OA hardly modifies the laser-to-THz conversion efficiency, except for coma, for which the THz yield drops by a factor $\approx 5$ due to the lower plasma density.

Figures~\ref{ConesFocLPP}(c-h) illustrate the associated angular THz spectra computed at the distances for which the growth of the THz energy plotted in Fig.~\ref{ConesFocLPP}(b) is maximum. The reference case [Fig.~\ref{ConesFocLPP}(c)] exhibits a conical emission at angles 2.5°~$< \Theta <$~4° and extends up to 60~THz.
The THz spectrum in the spherical case [Fig.~\ref{ConesFocLPP}(d)] appears similar. However, it presents a broadening up to 75~THz, which we attribute to strong nonlinearities associated with the local density peaks observed in Fig.~\ref{ConesFocLPP}(a). Astigmatism [Figs.~\ref{ConesFocLPP}(e,f)] and coma [Figs.~\ref{ConesFocLPP}(g,h)] display a reduction of the THz spectral width ($\nu < 45$~THz). In the former case, we observe that the radiated wave appears rather axisymmetric despite the asymmetric character of astigmatism. It presents a strong on-axis contribution at low frequencies ($\nu < 20-25$~THz) and a conical emission at $\Theta \approx 4$° in the upper part of the spectrum ($\nu > 25$~THz). In the coma case, the azimuthally-varying plasma shape renders the THz spectrum asymmetrical with different frequency bandwidths for positive and negative values of $\Theta$. This spectrum comprises on-axis emission for $\nu < 35 - 40$~THz and conical emission for $\nu > 40$~THz at angles between 2° and 3° in absolute value. 

\begin{figure}
	\centering
	\includegraphics[width=\linewidth]{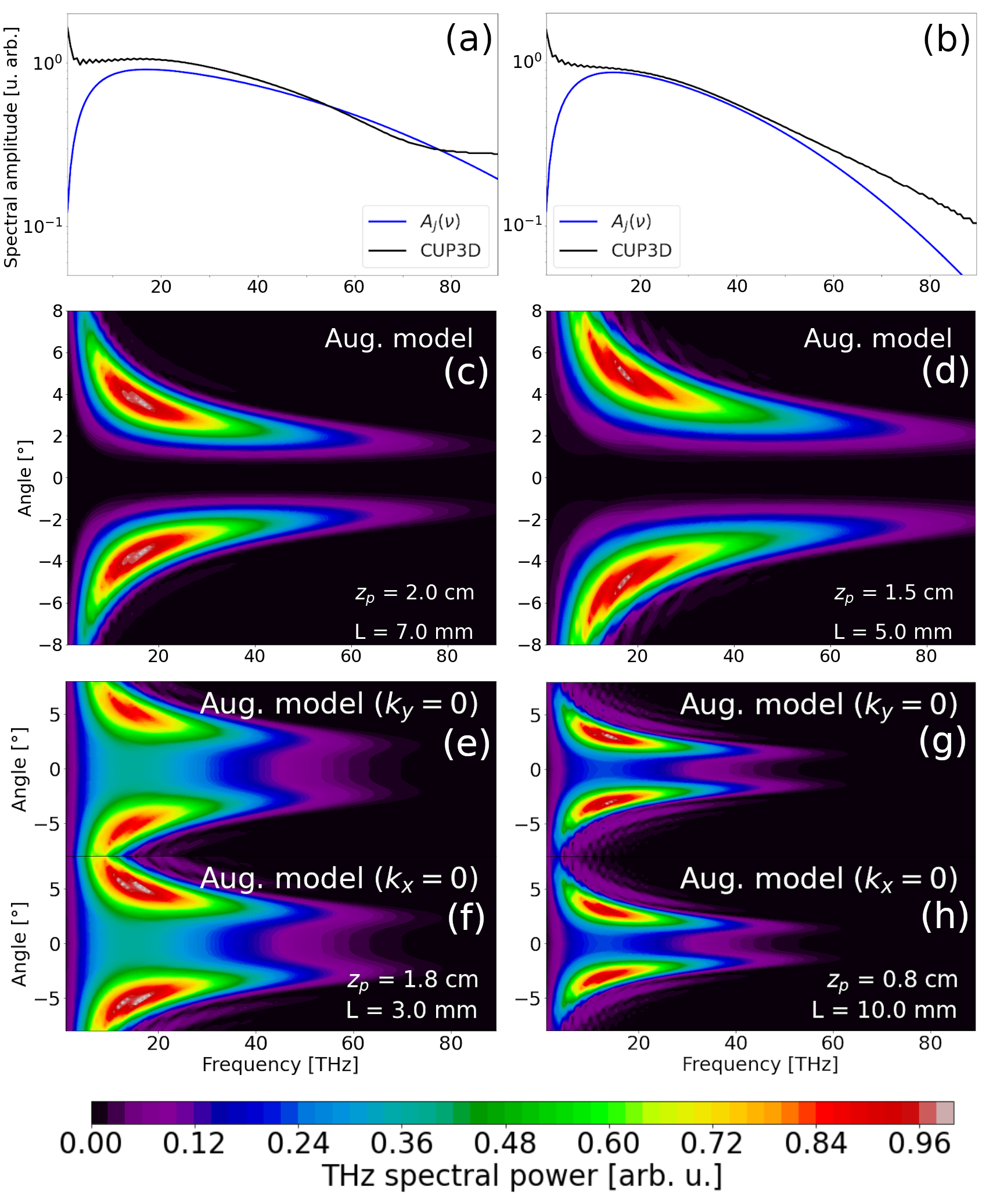}
	\caption{\label{dtJDMFocLPP} (a,b) Transversely integrated THz spectra of Fig.~\ref{ConesFocLPP} (black curves) and profiles $A_J(\omega)$ (blue curves) chosen to integrate our conical emission model (a) in the sphericity case and (b) the astigmatism case. (c-h) Angular THz spectra along (c,d,e,g) $k_y = 0$ and (f,h) $k_x = 0$ estimated from our augmented conical emission model for a laser propagation affected by (c) no OA (focus only), (d) sphericity, (e,f) astigmatism and (g,h) coma aberration. The parameters used are summarized in table \ref{tab:dtJFocLPP}.}
\end{figure}

For comparison, we integrated our augmented model to reproduce these spectra. The parameters used for its integration are summarized in table~\ref{tab:dtJFocLPP}.
We have selected our spectral profiles $A_J(\omega)$ so that they remain close to the transversely integrated UPPE THz spectra [see, e.g., solid blue curve in Fig.~\ref{FigFormeA}]. Here, only a plasma driver is accounted for in our test spectral profile. Introducing OA only shortens the bandwidth for asymmetric (e.g., astigmatism) aberrations, as exemplified by Figs.~\ref{dtJDMFocLPP}(a,b).

\begin{table}
\centering
   \begin{tabular}{| c | c | c | c | c | }
     \hline
      &\textbf{Focus} &\textbf{Sphericity}& \textbf{Astigmatism}& \textbf{Coma}\\ \hline
    $z_p$ [cm] & 2 & 1.5 & 1.8 & 0.8 \\ \hline
    $N_e^{\rm eff}$ [$10^{18}$ cm$^{-3}$] & 0.2 & 0.4 & 0.3 & 0.01 \\ \hline
    $L$ [mm] & 7 & 5 & 3 & 10 \\ \hline
    $l_d$ [mm] & 3.9 & 2.1 & 2.8 & 20 \\ \hline
   \end{tabular}
      \caption{Parameters used to integrate our augmented model for conical emission in the focused regime for LP-P laser pulses.}
      \label{tab:dtJFocLPP}
 \end{table} 
 
Figure~\ref{dtJDMFocLPP}(c) plots the angular THz spectrum estimated by our augmented model for the focus (reference) case. The distance $z_p$ reached by our linear pre-processor is chosen to match beam size and peak intensity similar to those developed by the UPPE runs. The predicted emission angles $\sim 2.5 - 3$° are in reasonable agreement with those of our UPPE simulations [see Fig.~\ref{ConesFocLPP}(c)]. The sphericity case, shown in Fig.~\ref{dtJDMFocLPP}(d), describes a conical emission very similar to that of the reference case and predicts angles of about $3 - 4$°. In agreement with the UPPE calculations, these two figures show that sphericity has little impact on the produced THz spectrum. For astigmatism [Figs.~\ref{dtJDMFocLPP}(e,f)], our model reproduces spectral narrowing and relevant on-axis components. However, it slightly overestimates the spectral asymmetries by amplifying more azimuthal variations (up to $\approx 9$\% in amplitude). These asymmetries are due to the terms $I_S^\pm(k_x, k_y)$, which ignore nonlinear propagation effects attached to Kerr self-focusing and plasma defocusing; the latter locally modify the spatiotemporal distributions of the laser pulse and the plasma channel along $z$. 

Nonetheless, the THz emission angles ($\Theta \approx 4$°) are correctly restored. In the case of coma [see Figs.~\ref{dtJDMFocLPP}(g,h)], asymmetries emerging in the UPPE spectra are partially reproduced. They are also associated with the asymmetric diffraction caused by the term $J_{\rm tr}(\vec{r}_\perp)$. Discrepancies with the UPPE simulation data can again be attributed to the nonlinear distortion of the plasma channel geometry omitted in our model. The overall spectral dynamics predicted by our model are nevertheless consistent with the UPPE simulation results, namely, a conical emission at angles $\Theta \approx 3$°, with a non-negligible on-axis contribution as $L \lesssim l_d$.

\subsection{CP-S Polarization}

We performed a similar study for circularly polarized laser pulses using the same laser-air parameters, except for the pump polarization state. 
We did not observe major modifications in the nonlinear laser fluency patterns calculated by UPPE for the four configurations of interest. We observed that the geometries of the laser beam and plasma remained similar to Fig.~\ref{E_NeLPP}, i.e., in the present focused geometry the laser polarization barely modifies pulse propagation in the presence of optical aberrations.

\begin{figure}
	\centering
        \includegraphics[width=\linewidth]{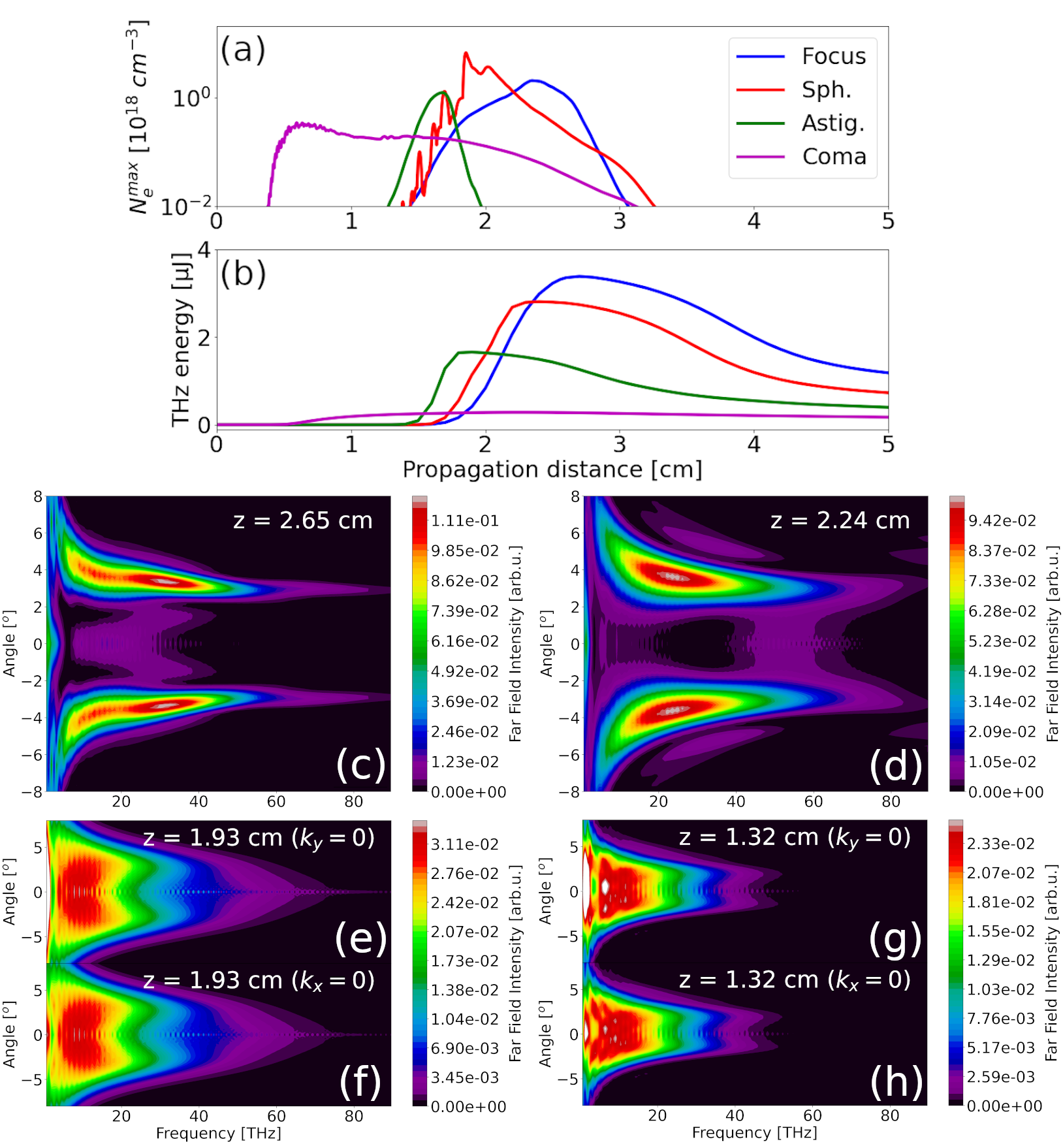}
	\caption{\label{ConesFocCPS} (a,b) Same quantities as in Fig.~\ref{ConesFocLPP} but for CP-S configuration. (c-h) Corresponding angular THz spectra calculated by UPPE (c,d,e,g) along $\vec{e_x}$ and (f,h) along $\vec{e_y}$ (c) in the reference case at $z = 2.65$ cm, (d) for sphericity at $z = 2.24$ cm, (e,f) for astigmatism at $z = 1.93$ cm and (g,h) for coma at $z = 1.32$ cm.}
\end{figure}

Figure \ref{ConesFocCPS}(a) illustrates the axial evolutions of the maximum electron densities reached without OA (focus, blue curve), with sphericity (red curve), astigmatism (green curve) and coma (magenta curve). Again, the peak plasma dynamics are very similar to those observed in the LP-P configuration. Figure~\ref{ConesFocCPS}(b) shows the related THz energy. We refind the well-known property following which CP-S laser pulses strongly increase the THz yield compared to their LP-P counterpart \cite{NJP}. Without OA (blue curve), the THz energy is enhanced by a factor $\approx 6$. This property holds with sphericity (red curve, $\approx \times 4$) and astigmatism (green curve, $\approx \times 3$), even if these two configurations significantly reduce the laser-to-THz conversion efficiency compared to the reference case. Coma aberration (magenta curve) is again harmful for THz radiation due to its detrimental action on the generated electron density.

Figures \ref{ConesFocCPS}(c-h) illustrate the associated angular THz spectra. These are close to the LP-P case, although with slightly narrower frequency bandwidth, which we attribute to relatively weaker nonlinearities. The action of astigmatism and coma on the THz distribution [see Figs.~\ref{ConesFocCPS}(e,f,g,h)] is similar to the LP-P case, although with a less pronounced asymmetry at negative angles in the coma case. Hence, LP-P and CP-S laser beams provide comparable THz emission patterns.

Let us now integrate our conical emission model. We assume that the CP-S polarization state barely modifies the parameters $L$ and $N_e^{\rm eff}$ so that the values presented in Table \ref{tab:dtJFocLPP} can be reused here. The $A_J(\omega)$ factors were checked to remain similar to those used in the LPP configuration.

\begin{figure}
	\centering
    \includegraphics[width=\linewidth]{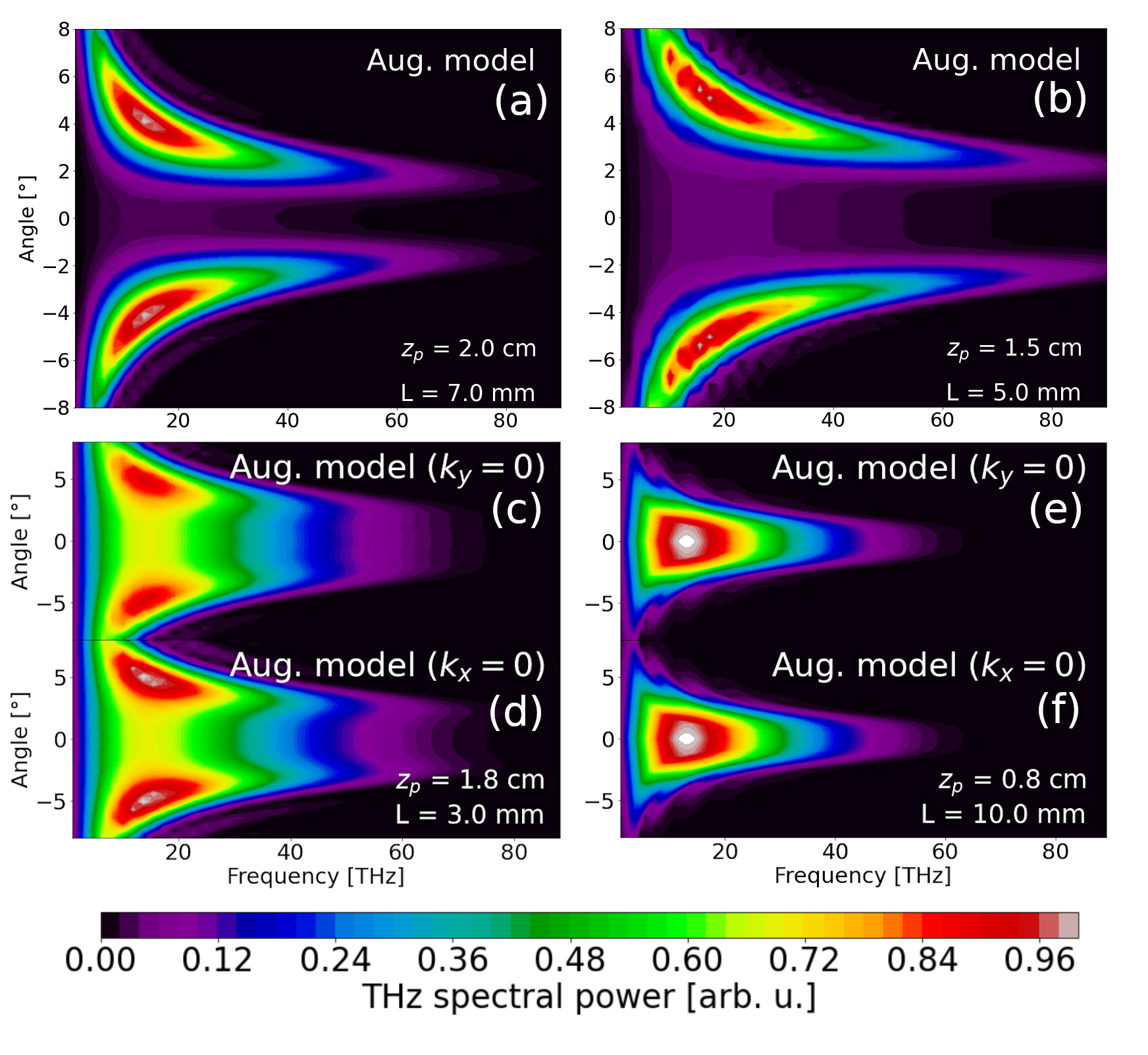}
	\caption{\label{dtJDMFocCPS} Same quantities as in Fig.~\ref{dtJDMFocLPP}(c-h), but for a CP-S laser pulse.}
\end{figure}

Figure~\ref{dtJDMFocCPS} plots the results of our model. Angular spectra appear very close to those calculated in the LP-P case, confirming again that laser polarization has a limited impact on conical emission. The overall angular spectra for asymmetric aberrations reasonably agree with their UPPE counterparts in the predicted spectral extent and production of on-axis components. We observe, however, a few discrepancies in the on-axis contributions, which are attributed to a strong dependence of the THz spectrum on the phase $\phi(\vec{r}_\perp)$ when $L \approx l_d$. We again emphasize that because the nonlinear responses of the medium alter the laser pulse during propagation, our model can overestimate some asymmetries in the THz spectra, e.g., for astigmatism [see Figs.~\ref {dtJDMFocCPS}(c,d)]. These asymmetries disappear in the coma case [Figs.~\ref{dtJDMFocCPS}(e,f)] that amplifies on-axis contribution and agrees better with the UPPE spectra. 

\subsection{Influence of optical aberrations in filamentation regime}

To explore THz emission in the filamentation regime modified by adaptative optics, we performed UPPE simulations of LP-P laser pulses without OA (focus), with sphericity ($c_4^0 = 10$~nm), astigmatism ($c_2^2 = 150$~nm) and coma ($c_3^{-1} = 100$~nm). Our 34-fs, two-color laser pulses now have a 2.5~mm initial radius and 1.5~mJ energy, and they are loosely focused with $f = 2.6$~m (equivalent to $c_2^0 = 300$~nm) in air. These simulations required using the linear pre-processor to propagate the laser pulses up to the threshold intensity of the first nonlinearities before using them as input data for the UPPE code. Laser beams without OA, with sphericity, astigmatism, and coma, were propagated over 2.3~m, 1.7~m, 1.7~m, and 1.5~m, respectively. The value of the distance $z$ where nonlinear propagation begins was always set to $z = 0$ by convention.

\begin{figure}
	\centering
	\includegraphics[width=\linewidth]{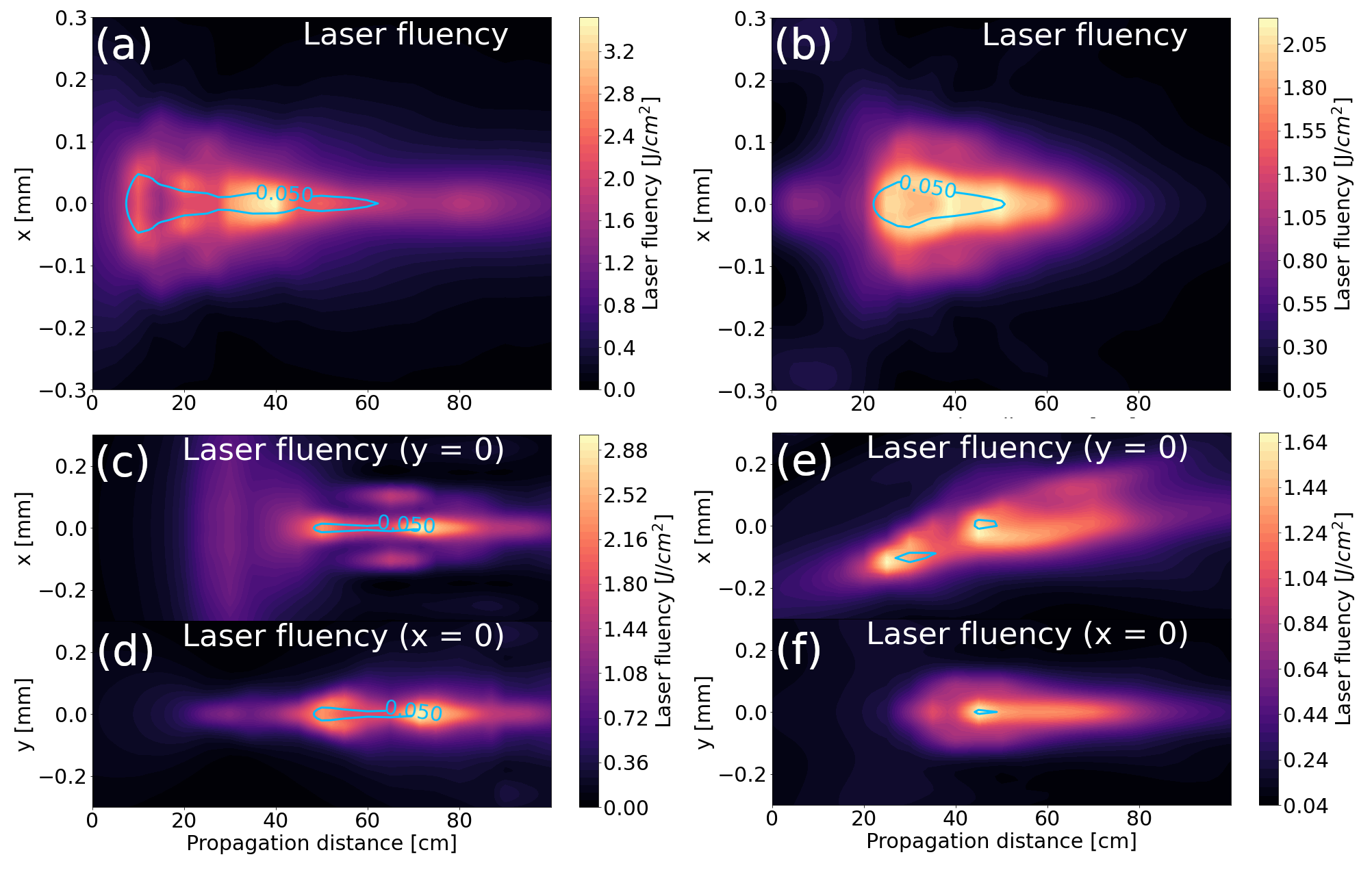}
	\caption{\label{E_NeFil} Laser fluency maps calculated in filamentation regime by the UPPE code (a,b,c,e) in the ($x$, $z$) plane and (d,f) in the ($y$, $z$) plane (a) in the reference (focus) case, (b) with sphericity, (c,d) astigmatism and (e,f) coma. The blue iso-contours represent the iso-densities $N_e^{\rm max} = 5 \times 10^{16}$~cm$^{-3}$.}
\end{figure}

Figure \ref{E_NeFil} shows the laser fluency calculated with UPPE. The reference case [Fig.~\ref{E_NeFil}(a)] displays a plasma zone that extends over several tens of centimeters. Kerr self-focusing/plasma defocusing cycles induce deviations from the patterns shown in Fig.~\ref{E_NeLPP}, particularly for pulses affected by asymmetric aberrations. The hollow zones observed in the reference case and with sphericity [Fig.~\ref{E_NeFil}(a,b)] at $z \approx 15$~cm and $z \approx 30$~cm, respectively, indicate a replenishment dynamics \cite{Mlejnek}. Unlike the focused regime, despite the presence of astigmatism over a long propagation range, a rather axisymmetric filament develops over the propagation axis. In addition, Figs.~\ref{E_NeFil}(e,f) for coma contain fewer hot spots compared to the fluency patterns observed in Figs.~\ref{E_NeLPP}(e,f). We conclude that in the filamentation regime, the propagation dynamics is dominated by the nonlinear balance between Kerr self-focusing and plasma defocusing, and linear effects due to OA are less important. 

\begin{figure}
	\centering
	\includegraphics[width=\linewidth]{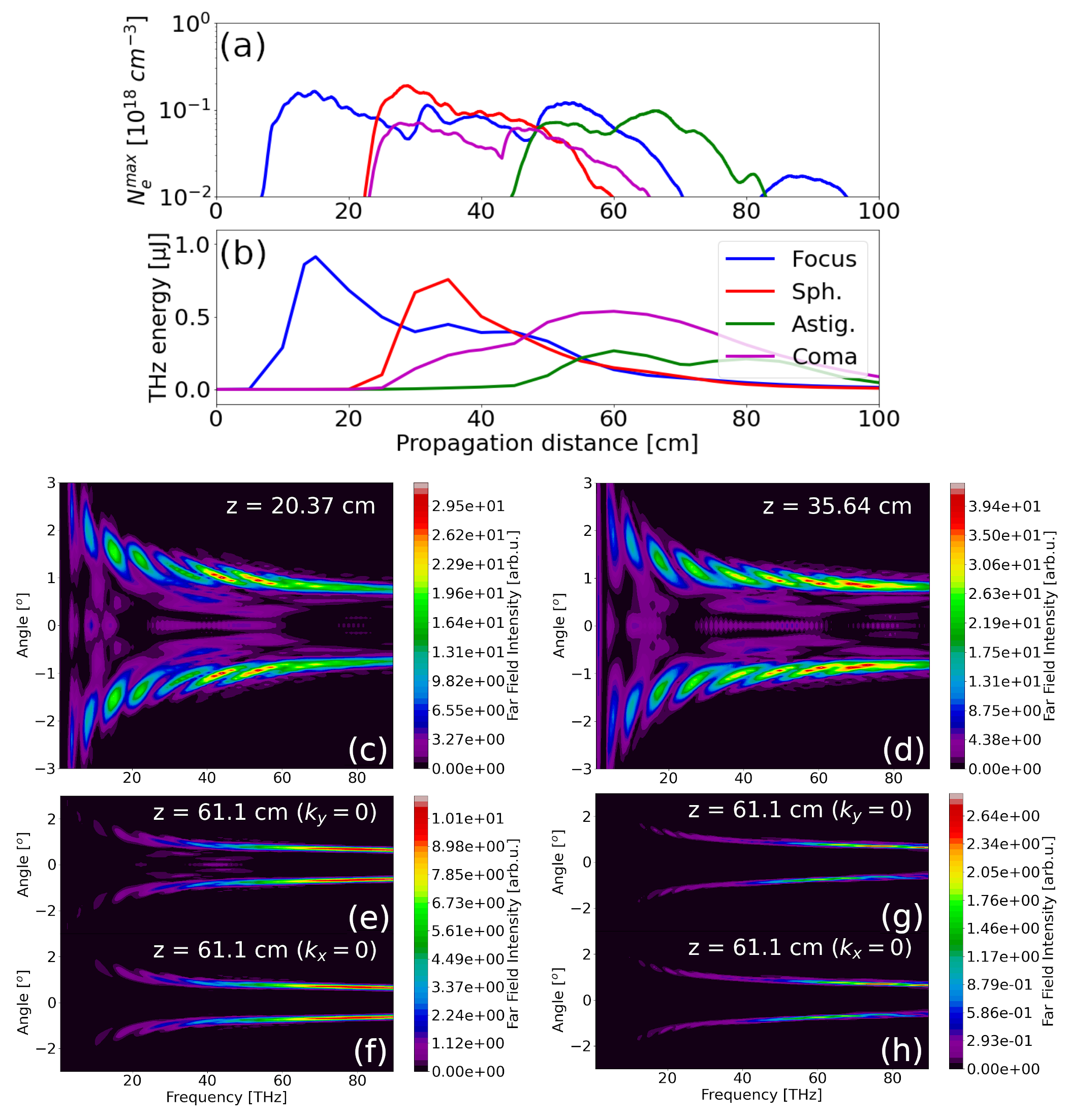}
	\caption{\label{ConesFilLPP} Top: (a) Maximum electron density and (b) THz energy ($\nu < 90$~THz) along $z$ in the reference case (focus, blue curve), with sphericity (red curve), astigmatism (green curve) and coma (magenta curve). Bottom: Angular THz spectra calculated with UPPE (c,d,e,g) along $\vec{e_x}$ and (f,h) along $\vec{e_y}$ (c) in the reference case at $z = 20.4$~cm, (d) with sphericity at $z = 35.6$~cm, (e,f) astigmatism at $z = 61.1$~cm and (g,h) coma at $z = 61.1$~cm.}
\end{figure}

Figure \ref{ConesFilLPP}(a) describes the longitudinal evolutions of $N_e^{\rm max}$ in the reference case (blue curve), with sphericity (red curve), astigmatism (green curve) and coma (magenta curve). The differences in the plasma onset distances are mainly due to the selected final distances of the pre-processed linear propagation runs, which vary from one case to another. Focus and sphericity induce comparable maximum densities, although the latter reduces the plasma length. On the other hand, asymmetric aberrations reduce plasma density levels by a factor of $\approx 2$ and also decrease the length of the plasma zone. Figure~\ref{ConesFilLPP}(b) plots the produced THz energy ($\nu < 90$~THz). We observe that all aberrations reduce the production of THz yield. OAs are thus generally harmful for THz generation in the filamentation regime, similar to their impact in the focused regime.

UPPE angular THz spectra are presented in Fig.~\ref{ConesFilLPP}(c-h). 
In both the reference case [Fig.~\ref{ConesFilLPP}(c)] and with sphericity [Fig.~\ref{ConesFilLPP}(d)], the THz radiation exhibits a relatively flat spectrum up to 90 THz, which is typical of the filamentation regime. Additionally, the emission angle remains nearly constant over a wide frequency bandwidth. Astigmatism and coma illustrated by Figs.~\ref{ConesFilLPP}(e,f) and \ref{ConesFilLPP}(g,h) modify the spectrum with an almost zero contribution at low frequencies ($\nu < 15-20$~THz). Like in the focused regime, astigmatism does not induce any asymmetry in the THz intensity radiated in filamentation regime. Coma tends to slightly unbalance THz emission between positive and negative $\Theta$ because of the misalignment of the plasma channel with respect to the optical axis. The observed THz emission angles remain similar to those of the reference case in the filamentation regime; although affected by OA, the effective plasma density is weaker. Moreover, the plasma length in this regime always satisfies the condition $L \gg l_d$, ensuring perfectly conical emission in the far field. The oblique lobes particularly visible at low frequencies in the spectra of Figs.~\ref{ConesFilLPP}(c,d) are a numerical artifact due to a coarser resolution in $\Theta$ at low frequency $\nu$.

\begin{table}
\centering
   \begin{tabular}{| c | c | c | c | c | }
     \hline
      &\textbf{Focus} &\textbf{Sphericity}& \textbf{Astigmatism}& \textbf{Coma}\\ \hline
    $z_p$ [m] & 2.6 & 2 & 2.2 & 2 \\ \hline
    $N_e^{\rm eff}$ [$10^{18}$ cm$^{-3}$] & 0.01 & 0.01 & 0.01 & 0.01 \\ \hline
    $L$ [cm] & 5 & 5 & 5 & 5 \\ \hline
    $l_d$ [cm] & 2 & 2 & 2 & 2 \\ \hline
   \end{tabular}
      \caption{Parameters used to integrate our augmented conical emission model in filamentation regime.}
      \label{tab:dtJFil}
 \end{table}
 
\begin{figure}
	\centering
	\includegraphics[width=\linewidth]{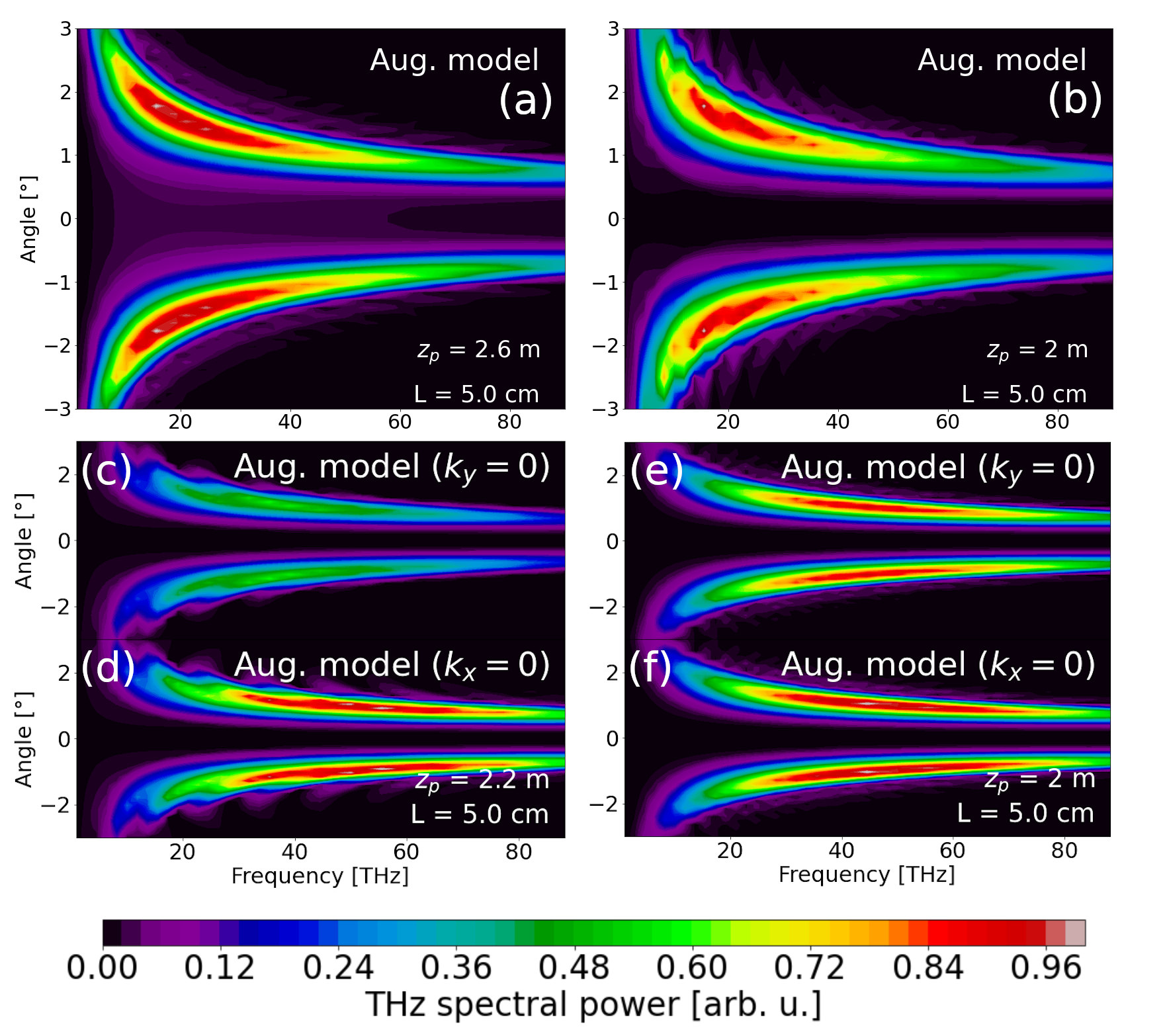}
	\caption{\label{dtJDMFilLPP} Angular THz spectra (a,b,c,e) along $\vec{e_x}$ and (d,f) along $\vec{e_y}$ estimated using our augmented conical emission model for a laser propagation affected by (a) no OA (lens of 2.6~m focal length), (b) sphericity, (c,d) astigmatism and (e,f) coma. Key parameters used for integration are detailed in table \ref{tab:dtJFil}.}
\end{figure}

To end with, we integrated our conical emission model using the parameters summarized in Table \ref{tab:dtJFil}. The spectral distribution $A_J(\omega)$ (dashed blue curve in Fig.~\ref{FigFormeA}) was chosen in each case to approach the broadband UPPE spectral profiles, which did not change significantly by adding OA. These include $32\%$ of Kerr contribution in amplitude ($10\%$ in energy ratio) extending till $\sim$70~THz. The results provided by our model are illustrated in Figs.~\ref{dtJDMFilLPP}. As in the UPPE simulations, OAs have no significant impact on the THz emission angles ($\Theta = 1-2$°), which are correctly restored here. Despite overestimated asymmetries in Figs.~\ref{dtJDMFilLPP}(c,d), our augmented model restores the vanishing of the low-frequency region for astigmatism and coma, together with the main trends of the UPPE spectrum.

\section{Conclusion}
\label{sec:conclusion}

In this paper, we have derived a so-called ''augmented'' conical emission model, able to satisfactorily reproduce angular THz spectra provided by 3D UPPE simulations, both in focused propagation and filamentation regimes. This model is called ''augmented'' because it still relies on the key parameters found by You et al.~\cite{You2012}. Yet, it includes complementary information on three other crucial quantities, namely, (i) the spectral distribution of the THz source in the nonlinear propagation regime of interest, (ii) the transverse profile and phase of the beam when it undergoes lensing effect or more complex phase distortions such as, e.g., those induced by deformable mirrors, and (iii) Kerr contribution to the overall THz radiation when self-focusing may become relevant, e.g., in the long-range filamentation regime. Our augmented model also addressed the role of pulse envelopes, which has been shown to be relatively limited, and it could be generalized to circularly polarized two-color pulses.

We carefully and systematically compared the results obtained from our augmented model with comprehensive UPPE simulation data. From this, we find an alternative to running heavy simulations before evaluating the angular spectra of the THz to interpret current experiments. Indeed, it appears sufficient to model a class of spectral profiles appropriately with respect to the targetted nonlinear propagation regime (tightly or loosely focused geometry) and appreciate the weight of the Kerr contribution to get reliable angular radiation patterns. Alternatively, LC-computed spectra could have been used, or, if available, the 3D UPPE spectral profiles could directly be implemented in the spectral amplitude $A(\omega)$. However, we have shown that simple test functions for the spectral distributions are sufficient to obtain valuable radiated intensities.

Importantly, instead of selecting one single radiated spectrum at a given propagation distance, we probed, for the first time to our knowledge, the whole spectral dynamics by examining changes in the radiated patterns for various plasma lengths around the dephasing length $l_d$. It appears that a good knowledge of the spectral dynamics along the propagation range covering the THz emitting zone is much more instructive than only analyzing the spectrum at the end of the plasma channel.

In connection with this property, our 3D UPPE simulations displayed evidence that the choice of the actual plasma length $L$ must be given by the length over which THz field emission takes place. This length is generally much shorter than the overall plasma extent along the optical axis. By computing the evolution of the relative phase between the two laser harmonics numerically along this effective plasma length (Fig.~\ref{FigTHzMaps}), we have proven that in focused propagation geometry $L$ remains close to $l_d$, which justifies why most of the radiated spectra measured in such emission regimes \cite{Andreeva,Ushakov,Nikolaeva} contain both on-axis emissions at small THz frequencies and net conical emission at higher frequencies. In contrast, a collimated or loosely focused propagation promotes plasma lengths $L \gg l_d$ and leads to conical emission in the entire THz frequency range.

In conclusion, we think that the present study provides the terahertz community with a deeper understanding of the THz spectral dynamics and additional tools to predict angular THz radiation patterns delivered by two-color air plasmas in future experiments.

\appendix

\section{THz emission model}
\label{app1}

Here, we establish a general conical emission model that includes laser envelope profiles in time and in the transverse plane. Let $t$ be the time variable and $\vec{r}$ be the position of the THz wave detector. The frame's origin is located at the center of the plasma with length $L$ assumed to be finite. In what follows, $\Theta$ corresponds to the polar angle at which the radiation is emitted with respect to the optical axis; $\Phi$ is the azimuthal angle.

Helmholtz's theorem gives the vector potential $\vec{A}(\vec{r}, t)$ emitted by a plasma source of volume $V$ with current density $\vec{J}(\vec{r}, t)$ as
\begin{equation}
\label{28}
{\vec A}({\vec r},t) = \frac{\mu_0}{4 \pi} \int_V d^3{\vec r}' \frac{{\vec J}({\vec r}',t')}{|{\vec r} - {\vec r}' |},
\end{equation}
where $t' = t - |{\vec r} - {\vec r}' |/c$ corresponds to the delayed time at which the detector receives the signal generated in $t$. The associated electric field $\vec{E}(\vec{r}, t)$, in the absence of a scalar potential, follows from
\begin{equation}
\label{29}
{\vec E}({\vec r},t) = - \frac{\mu_0}{4 \pi} \int_V d^3{\vec r}' \frac{\partial_t{\vec J}({\vec r}',t')}{|{\vec r} - {\vec r}' |},
\end{equation}
and the magnetic field from
\begin{equation}
\label{bfield}
{\vec B}({\vec r},t) = \frac{\mu_0}{4 \pi} \int_V d^3{\vec r}' \frac{\nabla \times{\vec J}({\vec r}',t')}{|{\vec r} - {\vec r}' |}.
\end{equation}
For the photocurrent mechanism considered here, it is safe to assume that the relevant current density is transverse and that no charge separation occurs ($\nabla\cdot \vec J=0$), which justifies the use of Eq.~(\ref{29}) to compute the radiated far-field power. We note that for the transition Cherenkov (TC) mechanism \cite{Damico, DamicoNJP, Balakin, Penano2010, Clerici, ThielePRA}, where the relevant current density is longitudinal and $\nabla\cdot \vec J \neq 0$ produces a charge density, one would need to account for a scalar potential in Eq.~(\ref{29}) or resort to Eq.~(\ref{bfield}).

The temporal Fourier transform of Eq.~(\ref{29}) leads to
\begin{equation}
\label{30}
\widehat{{\vec E}}({\vec r},\omega) = \frac{i \omega \mu_0}{4 \pi} \int_V d^3{\vec r}' \frac{\mbox{e}^{i |{\vec r} - {\vec r}' | \omega/c}}{|{\vec r} - {\vec r}' |} \widehat{{\vec J}}({\vec r}',\omega)\,.
\end{equation}
We assume that the distance between the plasma and the detector is large compared to the filament dimensions, that is, $|\vec{r}| \gg |\vec{r'}|$. Cartesian, cylindrical and spherical coordinates are written ($x$, $y$, $z$), ($r$, $\varphi$, $z$) and ($r$, $\Theta$, $ \Phi$). The notation $'$ applies to the local coordinates in the plasma. The far-field assumption leads to the simplifications
\begin{equation}
\label{31}
|{\vec r} - {\vec r}' | \simeq r - \frac{{\vec r}' \cdot {\vec r}}{r} = r - \frac{c}{\omega} {\vec k}_\perp \cdot {\vec r'}_\perp - z' \cos \Theta
\end{equation}
with ${\vec k}_\perp = (\omega/c)(\sin\Theta \cos \Phi, \sin\Theta \sin \Phi)$, providing Eq.~(\ref{30bis_2}) of the main article. 

To evaluate the source terms $\vec{J}_e(\vec{r}, t)$ and $\vec{P}_K(\vec{r}, t)$ in Eq.~(\ref{JK}) for a two-color laser pulse, we employ the general expression of the LP-P two-color laser field [Eq.~(\ref{35_2})] and apply the assumptions discussed in the main text. The free-electron contribution $\vec{J}_e(\vec{r}, t)$ is calculated using the toy ionization rate and assuming a small ionization yield,
\begin{multline}\label{WSimp} W(E) = C E^2(t),\quad N_e(t) = C N_a \int_{-\infty}^t E^2(t') dt'.
\end{multline}
The square of the LP-P two-color laser field Eq.~(\ref{35_2}) with the assumptions discussed in the main text reads
\begin{multline}
\label{38_01}
E_L^2({\vec r},t) = \mathcal{E}_1^2 \cos^2{[\omega_0 t'_1 + \phi_1}] + \mathcal{E}_2^2 \cos^2{[2\omega_0 t'_2 + \phi_2]} \\
+ 2 \mathcal{E}_1 \mathcal{E}_2 \cos{[2\omega_0 t'_2 + \phi_2]} \cos{[\omega_0 t'_1 + \phi_1]} ,
\end{multline}
where we omitted the arguments of $\mathcal{E}_1(\vec{r}_\perp, t'_1)$, $\mathcal{E}_2 (\vec{r}_\perp, t'_2)$, $\phi_1({\vec r}_\perp)$, and $\phi_2({\vec r}_\perp)$ for conciseness.
For the plasma term, straightforward integration of Eq.~(\ref{WSimp}) leads to linear and sine temporal dependencies in time when slowly varying envelopes are assumed. The product $N_e(t) {\vec E}_L(t)$ thus contains quasi-stationary contributions in the terms
$$
\frac{CN_a}{4\omega_0} \mathcal{E}_1^2\sin[2\omega_0 t'_1 + 2\phi_1] \mathcal{E}_2 \cos[2\omega_0 t'_2 + \phi_2]\vec{e_x}$$
and 
$$
\frac{CN_a}{\omega_0}\mathcal{E}_1 \mathcal{E}_2 \sin[\omega_0 (2 t'_2 - t'_1) + \phi_2 -\phi_1] \mathcal{E}_1 \cos[\omega_0 t'_1 + \phi_1]\vec{e_x}.
$$
The treatment of the Kerr term proceeds similarly from Eq.~(\ref{PKerr}) without integration in time and, therefore, only contains cosine contributions. After retaining only the quasi-stationary terms, the two potential THz sources can then be expressed as
\begin{multline}
\label{38}
(\partial_t + \nu_c) \vec{J}_{ e, {\rm THz}}({\vec r}, t) = {J}_{e, 0}\ \mathcal{E}_1^2({\vec r}_\perp, t_1') \mathcal{E}_2({\vec r}_\perp, t_2') \\
\times \sin{\left[\Delta k z + \phi(\vec{r}_\perp) \right]} \vec{e_x}
\end{multline}
using ${J}_{e, 0} = 3 C N_a e^2/(8 m_e \omega_0)$, and
\begin{multline}
\label{37} 
{\vec P}_{K,{\rm THz}}({\vec r}, t) = P_{K,0}\ \mathcal{E}_1^2 ({\vec r}_\perp, t_1')  \mathcal{E}_2({\vec r}_\perp, t_2') \\
\times \cos{\left[\Delta k z + \phi(\vec{r}_\perp) \right]} \vec{e_x}
\end{multline}
with $P_{K,0} = 3 \varepsilon_0 \chi^{(3)}/4$, $\Delta k = \pi/l_d$, and $l_d$ denoting the dephasing length given in Eq.~(\ref{eq:ld}). With the assumption of separable laser envelopes in space and time [Eq.~(\ref{40_2})] we can simplify the relations (\ref{38}) and (\ref{37}), and the unified source term $\widehat{\vec{J}}(\vec{r} ,\omega)$, cf.\ Eq.~(\ref{JK}), reads:
\begin{equation}
\widehat{\vec{J}}(\vec{r} ,\omega) = \widehat{\vec{J}_0}(\vec{r}, \omega) J_{\rm tr}(\vec{r}_\perp) \mathcal{F}_t\!\left[{\cal{E}}_{ax, 1}^2(t_1') {\cal{E}}_{ax, 2}(t_2')\right]\!(\omega),
\end{equation}
where
$
J_{\rm tr}(\vec{r}_\perp) = \mathcal{E}_{\rm tr, 1}^2({\vec r}_\perp) \mathcal{E}_{\rm tr, 2}({\vec r}_\perp)
$
[see Eq.~(\ref{Jtr})], and 
\begin{multline}\label{J0}
\widehat{\vec{J}_0}(\vec{r}_\perp, z, \omega) = \frac{ J_{e, 0}}{\nu_c - i \omega} \sin{(\Delta k z + \phi(\vec{r}_\perp))} \vec{e_x} \\- i \omega P_{K,0} \cos{(\Delta k z + \phi(\vec{r}_\perp))} \vec{e_x}.
\end{multline}
$\mathcal{F}_t$ refers to Fourier transform in time, and using $t_{j = \{1,2\}}' = t - zn(j\omega_0)/c$ we can rewrite
\begin{align*}
& \quad {\cal F}_t\! \left[{\cal{E}}_{ax, 1}^2(t_1') {\cal{E}}_{ax, 2}(t_2') \right]\!(\omega) = \int {\cal{E}}_{ax, 1}^2(t_1') {\cal{E}}_{ax, 2}(t_2') e^{i\omega t} dt \\
& = e^{i k n(\omega_0) z}\int {\cal{E}}_{ax, 1}^2(t_1') {\cal{E}}_{ax, 2}\!\left(t_1' - \frac{\pi z}{ 2\omega_0 l_d}\right) e^{i\omega t_1'} dt_1', 
\end{align*}
with $k = \omega/c$ being the THz wavenumber in vacuum. We can use the convolution theorem to rewrite the integral as
\begin{align*}
& \quad \int {\cal{E}}_{ax, 1}^2(t) {\cal{E}}_{ax, 2}(t - \gamma) e^{i\omega t} dt \\
& = \frac{1}{2\pi} \iiint {\cal{E}}_{ax, 1}^2(t) e^{i\omega' t} {\cal{E}}_{ax, 2}(t'-\gamma) e^{i(\omega' - \omega) t'} dt dt' d\omega' \\ 
& = \frac{1}{2\pi} \iiint {\cal{E}}_{ax, 1}^2(t) e^{i\omega' t} {\cal{E}}_{ax, 2}(\tilde{t}) e^{i(\omega' - \omega) (\tilde{t}+\gamma)} dt d\tilde{t} d\omega' \\ 
& = \frac{e^{-i\omega\gamma}}{2\pi} \int {\cal F}_t\! \left[{\cal{E}}_{ax, 1}^2(t)\right]\!(\omega') {\cal F}_t\! \left[{\cal{E}}_{ax, 2}({t})\right]\!(\omega' - \omega)e^{i\omega'\gamma}  d\omega'
\end{align*}
where $\gamma=\pi z / (2\omega_0 l_d)$. Provided that the spectra of the temporal envelopes are sufficiently narrow (not too short pulses), we can conclude that the integrand is nonzero for $\omega' \ll \omega_0$ only. Then, for plasma lengths of at most a few $l_d$ we can neglect the $z$-dependent term $\exp(i\omega'\gamma)$ under the integral. Because we are interested in THz frequencies only ($\omega \ll \omega_0$), with the same argument we can neglect the factor $\exp(-i\omega\gamma)$ in front of the integral. Thus, we have established that 
\begin{equation}
\quad {\cal F}_t\! \left[{\cal{E}}_{ax, 1}^2(t_1') {\cal{E}}_{ax, 2}(t_2') \right]\!(\omega) \approx e^{i k n(\omega_0) z} C_{\rm env}(\omega) 
\end{equation}
with
\begin{equation}\label{Com}
C_{\rm env}(\omega) = \int {\cal{E}}_{ax, 1}^2(t_1') {\cal{E}}_{ax, 2}(t_1') e^{i\omega t_1'} dt_1' . 
\end{equation}
While this expression is valid for arbitrary (slowly varying) pulse envelopes, it is instructive to inspect it for Gaussian envelopes:
\begin{equation}
\label{43}
{\cal{E}}_{ax, j}(t) \propto \mbox{e}^{- (t-t_c^j)^2/t_j^2}
\end{equation}
with $t_j$ being the duration of the $j^{th}$ color envelope and $t_c^j$ its temporal offset. After straightforward computations, we obtain a Gaussian power spectrum
\begin{equation}
C_{env}(\omega) \propto \frac{t_1t_2\sqrt{\pi}}{t_{eq}} e^{- \frac{2 \Delta t^2}{t_{eq}^2}} e^{i \omega (t_c^1-\Delta t \frac{t_1^2}{t_{eq}^2}) } e^{- \frac{t_1^2 t_2^2}{4 t_{eq}^2} \omega^2},
\end{equation}
where $t_{eq} = \sqrt{2 t_2^2 + t_1^2}$ and $\Delta t=t_c^1-t_c^2$ is the temporal separation of the two envelopes. Most notably, the amplitude strongly decreases with increasing $\Delta t^2$. Interestingly, deviations from the above slowly-varying assumptions and $\omega \ll \omega_0$ would produce additional contributions of orders $\Delta t L/(\omega_0 t_{eq}^2 l_d)$ and $\omega \lambda_{\rm THz}/(\omega_0 l_d)$, respectively, in the argument of the sinc function of Eq.~(\ref{Kappa}). Deviations from the above requirement $\omega \gamma \ll 1$ would introduce a $z$-dependent correction $e^{- i \gamma \omega t_1^2/t_{eq}^2}$ into $C_{env}$. For a $0.8\,\mu$m fundamental pulse, this contribution remains small for THz frequencies $< 60$ THz (120 THz) as long as $L \leq 4\,l_d$ (resp. $L \leq 2\,l_d$).

The source term $\widehat{\vec{J}}(\vec{r}, \omega)$ hence reads
\begin{equation}\label{JSimp}
\widehat{\vec{J}}(\vec{r}, \omega) \approx \widehat{\vec{J}_0}(\vec{r}, \omega) J_{\rm tr}(\vec{r}_\perp) e^{i k z n(\omega_0)} C_{\rm env}(\omega) .
\end{equation}
By injecting relations (\ref{JSimp}) into equation (\ref{30bis_2}), the radiated electric field takes the form:
\begin{equation}\label{ModeleAmp}
\widehat{\vec E}(r,\Theta, \Phi,\omega) = \frac{\mu_0}{4 \pi r} \mbox{e}^{i k r} \sum_{j = J, K} C_j(\Theta, \Phi, \omega),
\end{equation}
where 
\begin{equation}
    \label{CJ}
    C_{J, K}(\Theta, \Phi, \omega) = A_{J, K}(\omega) \vec{I}_{J, K}(\Theta, \Phi, \omega).
    \end{equation}
Here, $A_{J, K}(\omega)$ are the THz spectral profiles associated with the photocurrents and Kerr response,
\begin{align}
A_J(\omega) &= i \omega C_{env}(\omega) \frac{J_{e, 0}}{\nu_c - i\omega}, \\
A_K(\omega) &= \omega^2 C_{env}(\omega) P_{K,0},
\end{align}
while $\vec{I}_{J,  K}(\Theta, \Phi, \omega)$ describes the interferences between the THz waves generated along the plasma channel:
\begin{align*}
\vec{I}_J(\Theta, \Phi, \omega) & = \int_{-\frac{L}{2}}^{\frac{L}{2}} dz' \mbox{e}^{i k z' [n(\omega_0) - \cos{\Theta}]} \\
& \times \frac{\mbox{e}^{i \Delta k z'} I_S^+(\Theta, \Phi,\omega) - \mbox{e}^{-i \Delta k z'} I_S^-(\Theta, \Phi,\omega)}{2 i} \vec{e_x}, \\
\vec{I}_K(\Theta, \Phi, \omega) & = \int_{-\frac{L}{2}}^{\frac{L}{2}} dz' \mbox{e}^{i k z' [n(\omega_0) - \cos{\Theta}]} \\
& \times \frac{\mbox{e}^{i \Delta k z'} I_S^+(\Theta, \Phi,\omega) + \mbox{e}^{-i \Delta k z'} I_S^-(\Theta, \Phi,\omega)}{2 } \vec{e_x}.
\end{align*}
The quantities $I_S^\pm(\Theta, \Phi, \omega)$ refer to the transverse integral contributions to the THz spectrum:
\begin{equation}
I_S^\pm(\Theta, \Phi,\omega) = \iint d^2 \vec{r}'_\perp J_{\rm tr}(\vec{r}_\perp') e^{\pm i \phi(\vec{r}_\perp')} \mbox{e}^{- i {\vec k}_\perp \cdot {\vec r}_\perp'},
\end{equation}
with ${\vec k}_\perp = (\omega/c) (\sin\Theta \cos \Phi, \sin\Theta \sin \Phi)$. The functions $I_S^\pm(\Theta, \Phi,\omega)$ can be rewritten as
\begin{equation}
I_S^\pm(\Theta, \Phi,\omega) = \mathcal{F}_\perp \left[J_{\rm tr}(\vec{r}_\perp') e^{\pm i \phi(\vec{r}_\perp')} \right](\Theta, \Phi,\omega),
\end{equation}
where $\mathcal{F_\perp}$ represents the transverse Fourier transform. 

It can be checked that the elementary transformations
\begin{align}
\label{LPPtoCPSJ} \sin[\Delta k z + \phi(\vec{r}_\perp')]\vec{e_x} &\to \binom{\sin[\Delta k z + \phi(\vec{r}_\perp')]}{\cos[\Delta k z + \phi(\vec{r}_\perp')]} \\
\label{LPPtoCPSPk} \cos[\Delta k z + \phi(\vec{r}_\perp')] \vec{e_x} &\to \binom{\cos[\Delta k z + \phi(\vec{r}_\perp')]}{-\sin[\Delta k z + \phi(\vec{r}_\perp')]}
\end{align}
allow one to treat similarly CP-S laser pulses, cf.\ Eq.~(\ref{ECPS}).

Straightforward calculation of the integrals $\vec{I}_{J, K}(\Theta, \Phi, \omega)$ gives, for a LP-P laser pulse:
\begin{align}
\label{IjSimpApp} \vec{I}_J^{\rm LP-P}(\Theta, \Phi, \omega) &= \frac{L}{2} \left[I_S^+ \kappa_+ - I_S^- \kappa_- \right] \vec{e_x}, \\
\label{INlSimpApp} \vec{I}_{K}^{\rm LP-P}(\Theta, \Phi, \omega) &= i \frac{L}{2} \left[I_S^+ \kappa_+ + I_S^- \kappa_- \right] \vec{e_x},
\end{align}
and for a CP-S pulse:
\begin{align}
\label{IjSimpCPSApp} \vec{I}_J^{\rm CP-S}(\Theta, \Phi, \omega) &= \frac{L}{2} \binom{I_S^+ \kappa_+ - I_S^- \kappa_-}{I_S^+ \kappa_+ + I_S^- \kappa_-}, \\
\label{INlSimpCPSApp} \vec{I}_{K}^{\rm CP-S}(\Theta, \Phi, \omega) &= i \frac{L}{2} \binom{I_S^+ \kappa_+ + I_S^- \kappa_-}{-I_S^+ \kappa_+ + I_S^- \kappa_-}.
\end{align}
The quantities $\kappa_\pm$ are defined by:
\begin{align}
\kappa_\pm &= \mathrm{sinc}(\alpha_\pm) \\
\alpha_\pm &= \frac{1}{2} k L \left(n_0 - \cos \Theta \pm \frac{\lambda_{\rm THz}}{2 l_d} \right),
\end{align}
with $\mathrm{sinc}(x) = \sin x / x$ and $k=\omega/c = 2\pi/\lambda_{\rm THz}$.

Equations (\ref{Com}), (\ref{IjSimpApp}) and (\ref{INlSimpApp}) [alternatively (\ref{IjSimpCPSApp}) and (\ref{INlSimpCPSApp})] can finally be injected into the relation (\ref{ModeleAmp}) to obtain the far-field spectral intensity:
\begin{equation}\label{ModeleInt}
\left|\widehat{\vec E}(r,\Theta, \Phi,\omega)\right|^2 = \frac{\mu_0^2}{16 \pi^2 r^2} \left|\sum_{j = J, K} C_j(\Theta, \Phi, \omega) \right|^2.
\end{equation}
We have implicitly assumed that the dephasing length $l_d$ is constant in space, although the source terms of THz radiation include radial variations. This means that similarly to You et \textit{al.}'s model, we need to select an effective electron density level $N_e^{\rm eff}$ to estimate $l_d$ and thus evaluate our model. 

Note that similarly to \cite{You2012}, we ignore plasma absorption and opacity in the THz frequency range. These assumptions can be justified as follows. By neglecting the Kerr contribution and focusing on plasma dispersion, Eq.~(\ref{n_plasma}) reads 
\begin{equation}
\begin{split}
    n^2(\omega) & = n^2_{\rm opt}(\omega) - \frac{\omega_p^2}{\omega^2(1 + i \nu_c/\omega)} \\
    & = n^2_{\rm opt}(\omega) - \frac{\omega_p^2}{\omega^2+\nu_c^2} + i \frac{\nu_c}{\omega}\frac{\omega_p^2}{\omega^2+\nu_c^2}.
    \end{split}
\end{equation}
The real and imaginary parts of the medium index are given by
\begin{align}
\label{Re_n}    \Re[n(\omega)] & = \left(\frac{|n^2(\omega)|+\Re [n^2(\omega)]}{2} \right)^{1/2}, \\
\label{Im_n}    \Im[n(\omega)] & = \left(\frac{|n^2(\omega)|-\Re [n^2(\omega)]}{2} \right)^{1/2}.
\end{align}
For $\Re [n^2(\omega)]>0$, attenuation of THz waves through $\Im[n(\omega)]$ is due to plasma absorption ($\nu_c\approx3$~THz), which recovers Kim et al.'s\cite{KimNP} approximation $\Im[n(\omega)] \approx \nu_c \omega_p^2/(2 \omega (\omega^2+\nu_c^2))$ in the limits $\nu_c/\omega \ll 1,\, n^2_{\rm opt}(\omega) \approx 1$. At density levels $N_e \lesssim 10^{17}$~cm$^{-3}$, the absorption length $L_a = c/(\omega \Im[n(\omega)])$ is of the order of $\sim 100$~µm at $\omega = 2\pi\nu_c$. As typical filaments have a radius of $a \approx 20$ µm and THz radiation is emitted at angles $\Theta \approx 5$°, the propagation distance of those waves to exit the plasma is $\approx 200$~µm $\sim L_a$, meaning that plasma absorption should already have a limited effect on the emitted THz field. At higher frequencies, $L_a$ quickly increases ($\sim 500$~µm at 5~THz), and absorption becomes negligible.

In the lower region of the THz spectrum, $\Re [n^2(\omega)]=n^2_{\rm opt}(\omega) - \omega_p^2/(\omega^2+\nu_c^2)$ may become negative. Then, the attenuation of the THz waves through $\Im[n(\omega)]$ is due to the plasma opacity. However, this effect barely impacts the forward emitted spectrum, as the THz field is emitted at the front of the plasma, i.e., before the plasma builds up \cite{Koehler2011,Dechard2017}, so that the local $N_e$ and thus $\omega_p$ take smaller values.

\section{Derivation of You et \textit{al.}'s model}
\label{app2}

The original You et \textit{al.}'s model [Eq.~(\ref{You})] \cite{You2012, You2013} can be easily derived from our augmented emission model. We assume a homogeneous cylindrical plasma of length $L$ and radius $a$. Thus, the source term $J_{\rm tr}$ contained in $I_S^\pm(\Theta, \Phi,\omega)$, is constant on the disk of radius $r_\perp < a$ and zero outside. Moreover, we introduce $\phi_0$ as the constant value taken by the phase $\phi(\vec{r}_\perp)$. Hence, we can write
\begin{align*}
I_S^\pm &= e^{\pm i \phi_0} \iint d^2 \vec{r}_\perp' e^{- i \vec{k}_\perp \cdot \vec{r}_\perp'} \\
&= e^{\pm i \phi_0} \int_0^a dr_\perp' r_\perp' \int_0^{2 \pi} d\varphi' e^{- i k_\perp r_\perp' \cos \varphi' } \\
&= 2 \pi e^{\pm i \phi_0} \int_0^a dr_\perp' r_\perp' J_0(k_\perp r_\perp'), \\
\end{align*}
with $J_0(x)$ being the Bessel function of first kind and rank 0. The simplifications continue by using the identity $\int dx x J_0(x) = x J_1 (x)$ \cite{Gradshteyn}, where $J_1(x)$ describes the Bessel function of first kind and rank 1:
\begin{equation}
I_S^\pm(\Theta, \omega) = 2 \pi e^{\pm i \phi_0} a^2 \frac{J_1(\beta)}{\beta},
\end{equation}
where we used $k_\perp = k \sin \Theta$ and $\beta = a k \sin \Theta$. 

If we now neglect the contribution of the Kerr response to the THz radiation ($A_{K}(\omega) = 0$), assume that the laser pulse is linearly polarized (LP-P) and $n(\omega_0)=1$, we can evaluate Eq.~(\ref{ModeleInt}) as
\begin{multline*}
\left|\widehat{E}(r,\Theta,\omega)\right|^2 = \frac{\mu_0^2}{16 r^2} a^4 L^2 \left|A_J(\omega)\right|^2 \left(\frac{J_1(\beta)}{\beta} \right)^2 \\ \times \left| \mbox{e}^{i \phi_0} \kappa_+ - \mbox{e}^{- i \phi_0} \kappa_- \right|^2,
\end{multline*}
where the term depending on $\alpha_\pm$ develops as:
\begin{equation}
\left| \mbox{e}^{i \phi_0} \kappa_+ - \mbox{e}^{- i \phi_0} \kappa_-  \right|^2 = \kappa_+^2 + \kappa_-^2 - 2 \kappa_+ \kappa_- \cos(2 \phi_0).
\end{equation}
The far-field THz intensity radiated by the plasma thus restores You \textit{et al.}'s Eq.~(\ref{You}), where $A(\omega) \equiv A_J(\omega)$.

\section*{Author's Contributions}
All authors contributed equally to this work.

\section*{Acknowledgments}

\section*{Data availability}
The data supporting this study's findings are available within the present article. Complementary data can be made available from the corresponding authors upon reasonable request.

\bibliography{references} 

\end{document}